\documentclass[a4paper,11pt, dvipsnames]{article}
\pdfoutput=1 
\DeclareUnicodeCharacter{2032}{\ensuremath{^{\prime}}}
\usepackage{jheppub} 

\usepackage{graphicx}
\usepackage{dcolumn}
\usepackage{bm}
\usepackage{bbm}
\usepackage[normalem]{ulem} 
\usepackage{tikz}
\usepackage{rotating}
\usetikzlibrary{decorations.pathmorphing}
\usetikzlibrary{decorations.markings}
\usetikzlibrary{calc}
\usepackage{verbatim}
\usepackage{mathrsfs}
\usepackage{mathtools}
\usepackage[dvipsnames]{xcolor} 
\usepackage{url}
\usepackage{tabularx}
\usepackage{multirow}
\usepackage{subcaption}
\hypersetup{colorlinks=true, citecolor=blue, urlcolor=RubineRed }
\usepackage[capitalise]{cleveref} 
\usepackage{fontawesome5} 

\definecolor{blue(munsell)}{rgb}{0.0, 0.5, 0.69}

\newcommand{\tr}{\text{Tr}}

\DeclarePairedDelimiter\vev{\langle}{\rangle}

\begin{document}

\title{Gravitational Waves from Confining Dark Sectors with Self-Consistent Effective Potentials}
\author{Rachel Houtz,}
\author{Martha Ulloa,}
\author{Mia West} 
\affiliation{Department of Physics, University of Florida, Gainesville, FL 32611, USA}

\emailAdd{rachel.houtz@ufl.edu}
\emailAdd{m.ulloacalzonzin@ufl.edu}
\emailAdd{miawest@ufl.edu} 

\abstract{

In this work, we present a self-consistent prediction for the gravitational wave signal arising from confinement-induced phase transitions in hidden non-Abelian $SU(N)$ gauge theories with $F$ light flavors. To do this, we impose perturbativity and unitarity constraints on the thermal effective potential to identify the portion of parameter space that admits a reliable effective field theory description. We also include the Polyakov-loop–improved finite-temperature potential for both $N=3$ and $N=4$, where $N$ is the number of dark colors, using an approximate computation of the mediating effects. We compute the resulting gravitational wave spectrum and delineate the regions of parameter space that remain phenomenologically viable after imposing theoretical consistency conditions. We find that these constraints make uncovering a stochastic background gravitational wave signal in this scenario more challenging, even for proposed future detectors.

}

\maketitle


\section{Introduction}

Gravitational waves (GWs) provide a powerful probe of physics beyond the Standard Model (SM). A stochastic GW background from a first order phase transition (FOPT) promises sensitivity to energy scales beyond the reach of current collider experiments~\cite{Baker:2019nia,LISA:2017pwj, Kosowsky:1992rz, Crowder:2005nr, Corbin:2005ny, Yagi:2011wg, Harry:2006fi, AEDGE:2019nxb, Seto:2001qf,
Punturo:2010zz,Caprini:2019egz,
LISACosmologyWorkingGroup:2022jok,
Caprini:2024hue,NANOGrav:2023hvm, EPTA:2023fyk}. Many extensions of the SM feature hidden sectors that undergo FOPTs. Among them, an intriguing possibility is a confining dark sector, in which a non-Abelian gauge group confines and can result in a FOPT. Models with confining gauge groups can naturally accommodate dark matter candidates and appear across a broad spectrum of hidden sector theories~\cite{Lonsdale:2017mzg, Boddy:2014yra, Hambye:2009fg, Han:2007ae, Baldes:2021aph, Croon:2019ugf, Howard:2021ohe}.  

In this work we study a confining $SU(N)$ gauge theory with $F$ fermions in its fundamental representation. In the SM, QCD is one such theory, though its confinement-induced chiral symmetry breaking phase transition is expected to be a crossover phase transition given the observed quark masses~\cite{Aoki:2006we,HotQCD:2018pds,Borsanyi:2020fev,Guenther:2022wcr}. Whether the FOPT appears even in the chiral limit of QCD is still up for debate, as lattice methods have yet to find a FOPT even for very light pion masses~\cite{Bazavov:2017xul, Kuramashi:2020meg, Dini:2021hug}. This may indicate that the quark masses necessary for a FOPT are smaller than naive estimates, or this may point towards a lack of a FOPT when $F=3$~\cite{Cuteri:2021ikv, Pisarski:2024esv}, though for $F>3$ a FOPT very well may exist below the conformal window~\cite{Hansen:2017pwe, 
Pisarski:1983ms, 
 Iwasaki:1995ij, Iwasaki:2003de, Nagai:2009ip, DeGrand:2015zxa, 
Engels:1988ph}. 

In a hidden sector, then, one can arrange combinations of $N$ and $F$ that could plausibly lead to a first order chiral symmetry breaking phase transition near confinement, providing a source for observable GWs~\cite{Belyaev:2025zse, 
Pisarski:1983ms, 
Brown:1990ev, Iwasaki:1995ij, Iwasaki:2003de, Ayyar:2025eim,  Nagai:2009ip, DeGrand:2015zxa, 
Engels:1988ph,  Croon:2024mde,  Agrawal:2025xul, 
Agrawal:2025wvf, Morgante:2022zvc, Schwaller:2015tja, Croon:2019iuh, Huang:2020crf, Helmboldt:2019pan, Reichert:2021cvs}. Characterizing this signal is challenging because low energy effective field theories (EFTs) of mesons, such as the Linear Sigma Model (LSM), lose predictive power when gluon interactions become significant near the phase transition. Still, the EFT that probes the chiral phase transition can be improved by incorporating gluonic confinement effects, using lattice data-informed coefficients where available~\cite{Helmboldt:2019pan, Reichert:2021cvs, Pasechnik:2023hwv, Fukushima:2003fw, Ratti:2005jh,  Huang:2020crf}. 

Following the strategy of~\cite{Pasechnik:2023hwv}, we construct the finite temperature effective potential starting from the LSM and then add continual improvements to obtain a prediction of the dynamics near the phase transition. We incorporate confinement effects using the Polyakov Loop Model (PLM)~\cite{Ratti:2005jh, Ratti:2006wg, Ratti:2007jf, Schaefer:2007pw, Kahara:2008yg, Schaefer:2008ax}, and thermal effects are calculated using the Cornwall-Jackiw-Tombouli (CJT) method~\cite{Cornwall:1974vz, Amelino-Camelia:1992qfe, Amelino-Camelia:1993rvt}. Linearly realized chiral theories like the LSM offer direct access to symmetry restoration at the phase transition, but they can extrapolate into a regime where nonperturbative confinement dynamics become important and are not fully captured. In pure Yang Mills or in the large-$N$ limit, analytic techniques like holographic models can be employed~\cite{Agrawal:2025xul, Agrawal:2025wvf, Morgante:2022zvc}. Related modeling frameworks used to study GWs from chiral symmetry breaking at confinement include the LSM-only and (Polyakov-loop improved) Nambu-Jona-Lasinio type constructions~\cite{Helmboldt:2019pan, Huang:2020crf, Schwaller:2015tja, Croon:2019iuh, Reichert:2021cvs}. 

In this work, we take the point of view that regardless of the choice of EFT and its shortcomings or advantages in capturing the properties of the phase transition, a basic requirement necessary to trust its predictions is self-consistency. This should hold even if coefficients or operators in the EFT are being informed by nonperturbative effects. 
We ensure self-consistency by demanding that the EFT is perturbative, unitary, and bounded from below in the region relevant for the phase transition. Previous work on GWs from confining phase transitions tends to sample from coupling values that are perturbative and unitary in the tree level potential; for example~\cite{Helmboldt:2019pan, Pasechnik:2023hwv} restrict the magnitude of the quartic couplings to be $<16\pi$. While this is a reasonable initial guide, it may not be sufficiently restrictive and could lead to an overestimation of the GW signal. Here we check perturbativity and unitarity throughout the temperature evolution of the thermal potential. We find that our requirements will significantly constrain the available couplings, and tend to weaken the strength of the resulting GW signal compared to naive expectations.

We further incorporate updated estimates of the bubble wall velocity, which influence the peak frequency and amplitude of the GW power spectrum~\cite{Ai:2024btx}.
We also include the Polyakov-loop improved potential and associated medium effects for both $N=3$ and $N=4$, enabling a more complete exploration of hidden sectors with $N=4$. In particular, recent proposals for $U(1)_A$-breaking operators in chiral EFTs depend explicitly on $N$~\cite{Csaki:2023yas}, making access to another $N$ benchmark important for examining the corresponding phase transitions.

This paper is organized as follows: in~\cref{ImproveEffPot}, we construct our finite-temperature effective potential, beginning with the LSM and incorporating the PLM, medium effects, and the CJT framework. In~\cref{EFTConsistency}, we examine the self-consistency of the EFT by imposing perturbativity, unitarity, and vacuum-stability constraints. In~\cref{sec:PT}, we outline our calculation of the GW signal, and we present our results in~\cref{sec:results}. We conclude in~\cref{sec:DiscussionConclusions} with a summary of our findings and a brief discussion of their implications.

\section{The Improved Effective Potential}
\label{ImproveEffPot}

We focus on a confining dark sector governed by an $SU(N)$ gauge group with $F$ quark flavors in the fundamental representation. We assume the dark sector has a global chiral symmetry structure analogous to SM QCD, i.e., an approximate $SU(F)_L \times SU(F)_R \to SU(F)_V$.\footnote{To be precise, the full chiral symmetry is an approximate $U(F)_L \times U(F)_R \to U(F)_V$, though we allow for large explicit breaking of the $U(1)_A$ symmetry.} We take as our starting point the LSM:
	\begin{align}
	\mathcal{L}_\text{LSM} 
		&= \text{Tr}\left[ \left(\partial_\mu\Phi\right)^\dagger\left(\partial^\mu\Phi\right)\right] 
			- V_\text{LSM}(\Phi)\,,
	\end{align}
where $\Phi$ is an $F\times F$ matrix that organizes the meson fields as
	\begin{align}
	\Phi_{ij} 
		&= \frac{1}{\sqrt{2F}} \left(\sigma + i\eta'\right)\delta_{ij} 
			+ \left(X_a + i\pi_a\right)T^a_{ij}\,.
	\end{align}
The $T^a$ are the $SU(F)$ generators and $\pi_a$ are the pseudo-Nambu Goldstone bosons (pNGBs) associated with spontaneous breaking of the $SU(F)_A$. The $X_a$ are the heavy bound states associated with unbroken $SU(F)_V$ generators, and the $\eta'$ is the pNGB associated with the explicitly broken $U(1)_A$ symmetry. The vev of the $\sigma$ field is identified with the chiral condensate, and so $\sigma$ can serve as the order parameter of the chiral symmetry breaking phase transition. The potential is
	\begin{align}
	V_{\text{LSM}}
		&= -m_\Phi^2\text{Tr}\left[\Phi^\dagger\Phi\right]
			+ \frac12 \left(\lambda_\sigma - \lambda_a\right)\text{Tr}\left[\Phi^\dagger\Phi\right]^2 
			+ \frac{F}{2}\lambda_a\text{Tr}\left[\Phi^\dagger\Phi\Phi^\dagger\Phi\right]
		\nonumber\\ 			
		&\qquad
			-2\left(2F\right)^{F/2-2}c\left(\text{det}\Phi + \text{det}\Phi^\dagger\right) \,.
	\label{eq:lsm}
	\end{align}
The $\det\Phi$ term preserves $SU(F)_A$ but explicitly breaks $U(1)_A$, and therefore directly contributes to the mass of the $\eta'$ pNGB. One could include additional explicit chiral symmetry breaking terms to Eq.~(\ref{eq:lsm}) above, which would generate masses for the $\pi^a$ fields as well. So long as we take these masses to be much smaller than the scales relevant for the dynamics of the phase transition,\footnote{Namely, the confinement scale $\Lambda$, the dynamical symmetry breaking scale $f_\pi$, and the temperature at which the phase transition occurs} we can neglect them in our analysis. We note, however, that it is easy to avoid massless particles in the dark sector by including small explicit breaking effects. 

The tree-level potential for the $\sigma$ field is
    \begin{align}
    V_0^{\rm LSM} (\sigma)
        &= - \frac12 m_\Phi^2 \sigma^2
            + \frac18 \lambda_\sigma \sigma^4
            - \frac{ c }{F^2} \sigma^F 
            \,.
    \end{align}
Above the chiral symmetry breaking scale, $\vev\sigma = 0 $ and below $\vev\sigma = f_\pi$, where we use $f_\pi$ to denote the dynamical scale of chiral symmetry breaking in the dark sector.

\subsection{The Polyakov Loop Improvement}
We expect a confining phase transition to occur near and influence the chiral symmetry breaking phase transition. We model this with the PLM, an EFT used to study $SU(N)$ confinement. The EFT is built from the gauge-invariant Polyakov loop operator:
	\begin{align}
	\ell(\vec x)
	   &= \frac1{N} \tr_c\, {\mathbf L} (\vec x) \,,
	\end{align}
where the trace runs over color space, and
	\begin{align}\label{eq:PolyakovMatrix}
	\mathbf{L}(\vec x)
		&= \mathcal{P} \exp \left[ i \displaystyle\int_0^{1/T}
			\! d \tau\,  A_4 (\vec x , \tau) \right]
	\end{align}
is the thermal Wilson line, $\mathcal{P}$ gives a path ordering, and the coordinates $\vec x,$ $\tau$ are the spatial directions and Euclidean time, respectively. 
The Polyakov loop $\ell$ is gauge invariant and its vev breaks the $Z_N$ center symmetry,\footnote{exact in pure $SU(N)$ Yang-Mills. In our case, the center symmetry is only approximate, explicitly broken by the presence of light fermions.} and so is a useful starting point from which to build operators in the EFT.  It also serves as the order parameter for the (de)confinement phase transition, with $\vev\ell \neq 0$ above the confining critical temperature, and $\vev\ell = 0$ below, recovering the $Z_N$ center symmetry. In the limit of zero chemical potential which is the relevant limit here, $\ell$ is real and thus $\ell=\bar{\ell}$.
For $N=3$, the form of the PLM potential is~\cite{Ratti:2005jh}:
	\begin{align}
    	V_{\rm PLM}^{(N=3)}
		&= T^4 \left[ - \dfrac{b_2(T)}{2} |\ell|^2
			- \dfrac{b_3}{6} (\ell^3 + \bar{\ell}^3) 
			+ \dfrac{b_4}{4} | \ell|^4 \right] \,,
        \label{eq:plm-3}
	\end{align}

where $b_2(T)$ is given by
	\begin{align}
	 b_2(T)
	 	&=a_0  + a_1\left(\frac{T_\Lambda}{T}\right) + a_2\left(\frac{T_\Lambda}{T}\right)^2 
			+ a_3\left(\frac{T_\Lambda}{T}\right)^3 \,.
	\end{align}
We use $T_\Lambda$ to denote the critical temperature of the \emph{confining} phase transition in order to differentiate it from $T_c$, the critical temperature of the chiral phase transition. The coefficients $a_{0, 1,2, 3}$ and $b_{3, 4}$ can be extracted from lattice studies, and their values are shown in Table~\ref{tab:PLMparams}. 
We expect that confinement triggers chiral symmetry breaking and therefore $\Lambda_{\rm conf} > f_\pi$, where $\Lambda_{\rm conf}$ 
is the confinement scale. We estimate that $T_\Lambda \sim \Lambda$ and therefore define the factor $\xi$ using
    \begin{align}
    T_\Lambda 
        &= \xi f_\pi\,, \quad \xi \geq 1 \,,
    \label{eq:xi}
    \end{align}
where we expect $\xi$ to be an $\mathcal{O}(1)$ number. We will eventually use $\xi = \{1, 2, 5\}$ as benchmarks when we present our results in ~\cref{sec:results}.

For  $N=4$, following \cite{Huang:2020crf}, the potential becomes
	\begin{align}
	V_{PLM}^{(N=4)}
		&= T^4 \left[ - \,\frac{ b_2(T) }2\, |\ell|^2
			+ b_4 |\ell|^4 
			+ b_6 |\ell|^6 \right] \,,
        \label{eq:plm-4}
	\end{align} 
and 
	\begin{align}
	b_2(T)
		& =a_0 
			+ a_1\left(\frac{T_\Lambda}{T}\right) 
			+ a_2\left(\frac{T_\Lambda}{T}\right)^2 
			+ a_3\left(\frac{T_\Lambda}{T}\right)^3 
			+  a_4\left(\frac{T_\Lambda}{T}\right)^4 \,.
	\end{align}
Again, the $a_{0, 1, 2, 3, 4}$, $b_{2, 4, 6}$ are extracted from lattice~\cite{Panero:2009tv}, see Table~\ref{tab:PLMparams}.
\begin{table}[h!]
	\centering
		\begin{tabular}{|c|c|c|c|c|c|c|c|c|}
		\hline
         & $a_0$ & $a_1$  & $a_2$ & $a_3$ & $a_4$ & $b_3$ & $b_4$ & $b_6$ \\
		\hline
		$N=3$ & 6.76 & -1.95 & 2.625 & -7.44 & -- & 0.75 & 7.5 & --	\\
		$N=4$ & 9.51 & -8.79 & 10.1 & -12.2 & 0.498 & -- & -2.46 & 3.23	\\
		\hline
		\end{tabular}
		\caption{The lattice-informed parameters for the PLM model,
        from \cite{Huang:2020crf}.}
		\label{tab:PLMparams}
	\end{table}
\subsection{Medium Effects}
With the Polyakov loop included, one can now connect the gauge and chiral sectors of the theory using $V_{\rm med}$. These thermal medium effects are given by:
    \begin{align}\label{VMedium}
    V_{\rm med}
        &= - 2F T \displaystyle\int 
            \frac{ d^3p}{(2\pi)^3}\, 
                \tr_c 
               \left( \, \ln \left[ 1 + \mathbf{L}_R e^{- E_p/T} \right] 
                    + \ln \left[ 1 + \mathbf{L}_R^\dagger e^{- E_p/T} \right] 
                    \right)\,,
    \end{align}
where $E_p=\sqrt{|\, \vec{p}\,|^2+m_q^2\,}$ is the energy of the quarks, and $\mathbf{L}_R$ is the Polyakov matrix in the quark representation $R$, which is fundamental for all cases considered here. As a result, we consider the same Polyakov operator as defined in \cref{eq:PolyakovMatrix}. The quark mass term we set to 
\begin{align*}
    m_q\sim g_q\sigma \,,
\end{align*}
where $g_q$ is an $\mathcal{O}(1)$ coupling constant. For the remainder of this work, we set $g_q=1$. In principle,  additional couplings between the mesons and Polyakov loops may exist (see e.g. \cite{Mocsy:2003qw}). 
However, we neglect them in this analysis.

\begin{figure}[h!]
\centering
\includegraphics[width=0.7\textwidth]{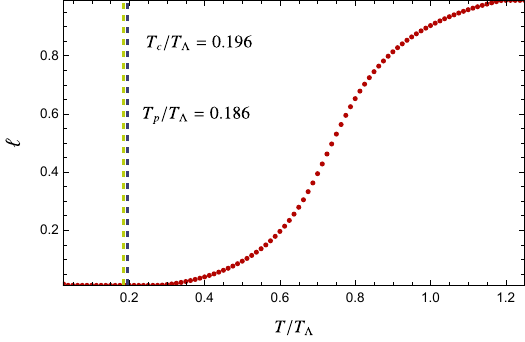}
\caption{ Polyakov loop for $F = 4$ and $N_c = 3$: $\ell$ as a function of $T$. 
As the temperature decreases, $\ell$ drops from a deconfined value 
$\ell\neq 0$ to $\ell \simeq 0$, and the steep fall 
signals the confining phase transition. The vertical dashed line in black indicates the ratio of the
critical temperature $T_c$ of the chiral phase transition to the critical temperature of the confining phase transition $T_\Lambda$, and in green the nucleation 
temperature $T_p$ to $T_\Lambda$, respectively. The benchmark point shown corresponds to 
$T_c = 833.21$ GeV, $T_p = 788.58$ GeV, $f_\pi = 2121.32$ GeV, and 
$T_\Lambda = \xi f_\pi$ with $\xi = 2$, together with 
$m_{\sigma}^2 = 1.0 \times 10^6\;(\text{GeV})^2$, $m_{\eta'}^2 = 1.54 \times 10^7\;(\text{GeV})^2$, and 
$m_{X}^2 = 5.8 \times 10^6\;(\text{GeV})^2$.
}
 
\label{fig:PolyakovLoopTransition}
\end{figure}

Taking the color traces over \cref{VMedium}, we use the usual $\tr_c\ln A=\ln\det A$ relation, coupled with the Cayley-Hamilton theorem to expand the determinant. 
For $N=3$, this is:
\begin{align}
    &\det\left[{1+\mathbf{L} e^{-E_p/T}}\right]
    \nonumber\\
    &\quad
    =\frac{1}{6}\Bigg[6+6\,\tr_c\, {\mathbf L}e^{-E_p/T}+\underbrace{\left[3\left[\tr_c\,{\mathbf L}\right]^2-3\tr_c\,{\mathbf L^2}\right]}_{\tr_c\,{\mathbf L^\dagger}}e^{-2E_p/T}+e^{-3E_p/T}\Bigg]\,,
\end{align}
where we make use of the reality condition of the Polyakov loop which holds in zero chemical potential, $\tr_c\,{\mathbf L^\dagger}=\tr_c\,{\mathbf L}$ to find
\begin{align}
    \ln\left[\det\left[1+\mathbf{L}_R e^{-E_p/T}\right]\right]=\ln\left[1+3\ell e^{-E_p/T}+3\ell e^{-2E_p/T}+e^{-3E_p/T}\right].
\end{align}
We also compute the $N=4$ case, where
\begin{align}
    &\det\left[{1+\mathbf{L}_R e^{-E_p/T}}\right]
    \nonumber\\
    &\quad=\frac{1}{24}\bigg[24+24\tr_c\, {\mathbf L}e^{-E_p/T}+\left[12\left[\tr_c\, {\mathbf L}\right]^2-12\tr_c\, {\mathbf L^2}\right]e^{-2E_p/T}
    \nonumber\\
    &\qquad+\underbrace{\left[4\left[\tr_c\, {\mathbf L}\right]^3-12\left[\tr_c\, {\mathbf L^2}\right]\tr_c\, {\mathbf L}+8\tr_c\, {\mathbf L^3}\right]}_{24\tr_c\, {\mathbf L^\dagger}}e^{-3E_p/T}+24e^{-4E_p/T}\bigg]\,.
\end{align}
We must now also introduce a charge-2 Polyakov loop $\ell_2$, \cite{Reichert:2021cvs}, through
\begin{align}
    \tr_c\, {\mathbf L^2}=N(\ell_2+\ell)\,,
\end{align}
which, following \cite{Pisarski:2001pe, Pereira:2023nqy}, we assume to be heavy and thus can be integrated out. Such an approximation should, however, be verified with the lattice. This same assumption is made in \cite{Huang:2020crf} to produce \cref{tab:PLMparams}.
Then once again applying the reality condition, the final expression is
\begin{align}
    \ln\left[\det\left[1+\mathbf{L}_R e^{-E_p/T}\right]\right]=\ln\left[1+\ell e^{-E_p/T}+\frac{3}{2}\ell^2e^{-2E_p/T}+\ell e^{-3E_p/T}+e^{-4E_p/T}\right].
\end{align}
The Polyakov loop parameter $\ell$ acts as an order parameter 
for the deconfining phase transition. However, we do not know exactly how it behaves across the phase transition and thus we employ the Mean Field Approximation (MFA) over $\ell$ to capture its effects \cite{Weise:2007cx}. This is expected to be a good approximation if the chiral phase transition does not significantly alter the dynamics of the deconfining phase transition. We assume this to be the case, and so for every $\sigma,T$ we minimize the potential with respect to $\ell$, which is itself constrained to lie between $\ell=0$ (confined) and $\ell=1$ (de-confined). In practice this means 
\begin{align}
    V_\text{PLM}(\sigma,T)+V_\text{med}(\sigma,T)=\underset{0\leq\ell\leq1}{\text{Min}}\left[V_\text{PLM}(\sigma,T,\ell)+V_\text{med}(\sigma,T,\ell)\right].
\end{align}
In \cref{fig:PolyakovLoopTransition} we show how the Polyakov loop parameter changes as a function of $T/\Lambda_\text{conf}$ in the MFA.

\subsection{The CJT Improvement}
\label{sec:CJT}

Here we derive the finite-temperature corrections to the LSM using the CJT formalism~\cite{Cornwall:1974vz, Amelino-Camelia:1992qfe, Amelino-Camelia:1993rvt}. This is based on an extended action that depends not only on the background field $\sigma$ but also on the expectation value of the two point Green's function. The CJT potential is given by 
\begin{align}
    V_{\mathrm{CJT}} 
        &= V_{0}\left( \sigma \right) 
            + \dfrac{1}{2} \sum_{i} \int_\beta \ln G_{i}^{-1} \left( \sigma, s, P \right) 
        \nonumber \\
        &\quad 
            + \dfrac{1}{2} \sum_{i}\int_\beta \left[ D^{-1} \left(  \sigma, s, P \right) G \left( \sigma, s, P \right) -1 \right] 
            + V_2 \left( \sigma, G\right) \,,
\end{align}
where \( V_{0}, \,  D^{-1}, \, V_2\), denotes the tree level potential, tree level propagator and the sum of the infinite sum of the two-particle irreducible (2PI) vacuum graphs. The scalar and pseudoscalar fields are denoted by \( s, P\), respectively, and $G$ is the dressed propagator.
Here we use the Hartree approximation~\cite{Hartree:1928, Fock:1930}, meaning, including only the one double-bubble diagram.
The effective potential of the improved LSM reads
\begin{align}
  V_{\mathrm{eff}}^{\mathrm{LSM}}(\sigma, T)
  = V_{0}^{\mathrm{LSM}}(\sigma)
  + V_{\mathrm{CJT}}^{\mathrm{LSM}}(\sigma, T)\,,
\end{align}
where the finite-temperature contribution is
\begin{align}
  V_{\mathrm{CJT}}^{\mathrm{LSM}}(\sigma, T)
  &= \frac{T^4}{2\pi^2} \sum_i
  \left[
    J_B\!\bigl(R_i^2\bigr)
    - \frac{1}{4}\bigl(R_i^2 - r_i^2\bigr)\,
      I_B\!\bigl(R_i^2\bigr)
  \right]\,.
  \label{eq:cjt}
\end{align}
The thermal functions used above are
\begin{align}
  J_B(R^2)
  &= \int_0^\infty \! dx\, x^2
     \ln\!\left(1 - e^{-\sqrt{x^2 + R^2}}\right), \\
  I_B(R^2)
  &= 2\,\frac{d J_B(R^2)}{d R^2}
   = \int_0^\infty \! dx\,
     \frac{x^2}{\sqrt{x^2 + R^2}}\,
     \frac{1}{e^{\sqrt{x^2 + R^2}} - 1}\,.
\end{align}
We introduced the dimensionless ratios
\begin{align}
  r_i \equiv \frac{m_i(\sigma)}{T}\,,
  \qquad
  R_i \equiv \frac{M_i(\sigma,T)}{T}\,,
\end{align}
where $m_i(\sigma)$ are the tree-level effective masses and
$M_i(\sigma,T)$ the thermally corrected masses. We compute these masses following the analysis of~\cite{Roder:2003uz}, which applied the CJT to chiral perturbation theory. We present our resulting expressions for the tree-level and dressed thermal masses~\cref{app:cjt}. Note that the CJT resummation in the Hartree-Fock approximation can be shown to be equivalent to Daisy and Super-Daisy resummation \cite{Amelino-Camelia:1992qfe,Amelino-Camelia:1993rvt}.

\subsection{The Full Improved Potential}

Pulling together all of the improvements above, our full effective potential is given by:
    \begin{align}
    V_{\rm eff} (\sigma, \ell, T)
        &= \underbrace{V_{\rm LSM}(\sigma)}_{\rm Eq.~\eqref{eq:lsm}}
            + \underbrace{V_{\rm PLM} (\ell, T)}_{\rm Eqs.~\eqref{eq:plm-3},\eqref{eq:plm-4}}
            + \underbrace{V_{\rm med} (\sigma, \ell, T)}_{\rm Eq.~\eqref{VMedium}}
            + \underbrace{V_{\rm CJT} (\sigma, T)}_{\rm Eq.~\eqref{eq:cjt}}
            \,,
    \end{align}
where $V_{\rm LSM}$ encodes the explict and spontaneous chiral symmetry breaking, $V_{\rm PLM}$ encodes the gluonic confining effects, $V_{\rm med}$ encodes the medium interactions between mesons and gluons, and $V_{\rm CJT}$ encodes the finite-$T$ corrections to the LSM. Also note that in the final expression, the Polyakov loop parameter $\ell$ is minimized over in the MFA.

\section{EFT Consistency}
\label{EFTConsistency}

We next examine the internal consistency of the EFT that governs the phase transition. Specifically, we require that the EFT remain under perturbative control when including thermal corrections, satisfy unitarity bounds, and possess a stable vacuum structure across the range of parameters relevant for the phase transition. The following subsections detail these three requirements and the quantitative criteria adopted in our analysis.

\subsection{Perturbativity}
\label{sec:pert}

It is well known that small bosonic masses at large enough $T$ result in a breakdown in the perturbative expansion of the thermal field theory, the so-called IR problem~\cite{Linde:1980ts, Arnold:1992rz}. In our setup, this problem appears when the $\sigma$ or $\pi^a$ effective masses get small.\footnote{In principle we can always lift the pion masses using explicit breaking effects. If the explicit breaking is small enough to justify our choice to ignore their impact on the phase transition, then the explicit breaking cannot at the same time be large enough to be relied upon to alleviate the IR problem.} 
Various resummation schemes exist to ameliorate the IR problem~\cite{Kajantie:1995dw, Croon:2020cgk, Croon:2020cgk, Gould:2019qek, Niemi:2018asa, Curtin:2016urg, Curtin:2022ovx}. Here we follow~\cite{Helmboldt:2019pan, Pasechnik:2023hwv} and use the CJT formalism as the phase transition studied here involves bound states. As noted above, the CJT resummation in the Hartree-Fock approximation is equivalent to Daisy and Super-Daisy resummation \cite{Amelino-Camelia:1992qfe,Amelino-Camelia:1993rvt}. 
It is therefore worthwhile to go beyond naive implementation of the CJT formalism by assessing its domain of validity. We note that we do not account for super-super-daisy resummation. The diagrams would pick up a factor of
    $\lambda_{\rm eff}=  { \lambda N_{\rm sp} T }/{M(\sigma, T)}$
when compared against the CJT result, where $N_{\rm sp}$ is the number of species running in the loop. When these diagrams are no longer subleading, the perturbative expansion is no longer trustworthy. We assess this with the heuristic that $\lambda_{\rm eff}$ is less than an optimistic loop factor:
    \begin{align}
    \lambda_{\rm eff}
        &=  \frac{ \lambda N_\text{sp} T }{M(\sigma, T)} < 16\pi \,,
    \end{align}
across over $\sim90\%$ of the potential between the metastable and true minima, at the percolation temperature. Note also that the Hartree-Fock approximation used here neglects 2PI diagrams with complicated topology (for example, diagrams with overlapping UV/IR divergences). Upon checking the contributions from naively leading neglected diagrams using~\cite{Curtin:2016urg}, we expect a similar or less constraining perturbativity bound. A full analysis of the neglected 2PI diagrams, however, is beyond the scope of our analysis here. 
By necessity, during the phase transition the tree-level mass of the $\sigma$ particle must cross through zero.\footnote{The Hartree-Fock approximation does not accurately account for this. See \cite{Roder:2003uz} for more details.} The dressed mass, however, typically receives a finite-$T$ contribution that shifts its mass with thermal corrections proportional to temperature and can ameliorate this effect. 
Despite this, the effective coupling can for some points still grow to be very large such that the perturbative expansion is no longer under control. If this occurs for a large fraction of the effective potential around the phase transition region, there is no way to trust the computation, and a more involved resummation would be required. As a heuristic for when this may occur, we give the effective couplings which are likely to be the largest:
\begin{align}
    \lambda_{\text{eff},\sigma}&=(3\lambda_\sigma-c_1) \frac{T}{\sqrt{|M_\sigma^2(\sigma,T)|}}\label{eq:lambdaeffsigma}\\
    \lambda_{\text{eff},\pi}  
        &=\left[ 
            (F^2+1)\lambda_\sigma
            +(F^2-4)\lambda_a
            -c_3 \right] \,
        \frac{T}{\sqrt{|M_\pi^2(\sigma,T)|}} \,.\label{eq:lambdaeffpi}
\end{align}

\begin{figure}[h!]
\centering
\includegraphics[width=0.9\textwidth]{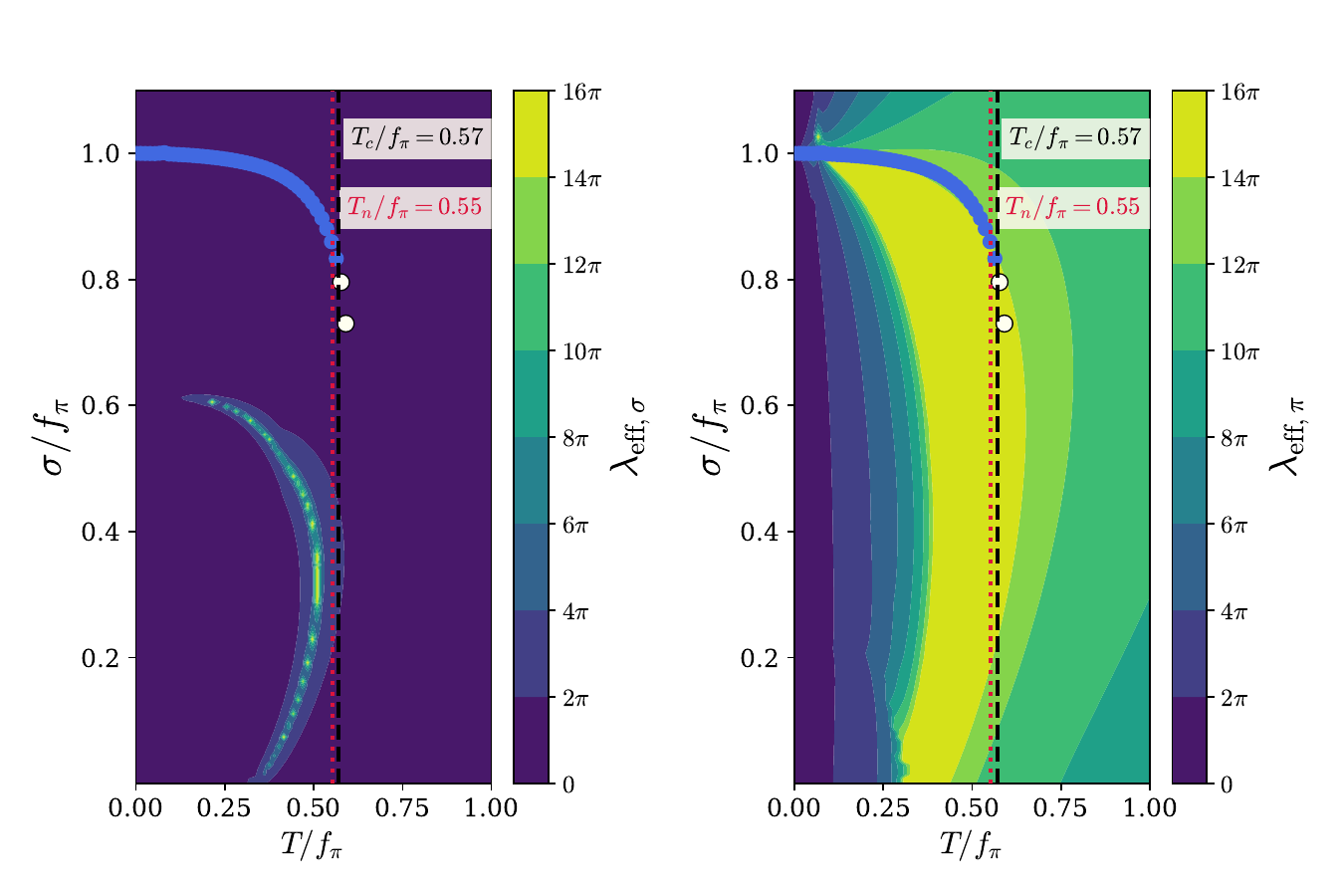}
\caption{ Contours of the effective couplings  $\lambda_{\text{eff},\sigma}$ and $\lambda_{\text{eff},\pi}$ defined in \cref{eq:lambdaeffsigma} and \cref{eq:lambdaeffpi} respectively, in the $(\sigma/f_\pi,\,T/f_\pi)$ plane. The benchmark shown corresponds to $N=3$, $F=3$, $T_c = 628.79~\text{GeV}$, $T_n = 608.82~\text{GeV}$, $f_\pi = 1102.27~\text{GeV}$, and $\xi = 1$. It is important to note that this point meets $49 \%$ of the perturbativity criteria. The dots indicate the position of the non-symmetric minimum as a function of temperature: blue dots denote that this minimum is metastable (false vacuum), while white dots denote that it is the true minimum at that temperature. Note the numerical noise associated with the convergence of the system of equations for the dressed masses when a particle mass passes through zero; see \cref{app:cjt} for details.
}

\label{fig:pert-plots}
\end{figure}

Here the coefficients $c_{1, 3}$ are given in Eq.~\eqref{eq:c1} and~\eqref{eq:c3}, which are used in the calculation of the thermal dressed masses in \cref{app:cjt}. We show and example of the evolution of $\lambda_{\rm eff}$ over the range of background $\sigma$ field values and as the temperature evolves in Fig.~\ref{fig:pert-plots}. From this, we see that $49\%$ of the thermal effective potential fails the heuristic that $\lambda_{\text{eff},\pi}<16\pi$ at the percolation temperature. Therefore, we class this point as unsafe.

\subsection{Unitarity}
\label{sec:unit}
A careful consideration of unitarity constraints is imperative to ensure consistency before considering GW
phenomenology, for example as detailed in \cite{Crawford:2024nun}. To place unitarity constraints on the tree-level LSM potential, we compute the $2\to2$ elastic scattering amplitudes of the meson fields in the broken phase from the tree-level diagrams in \cref{fig:DiagramTopologies}. We have listed in~\cref{app:4pt-feyn} the relevant Feynman rules. Using the results in \cite{Goodsell:2018tti}, we compute the zeroth partial wave for each scattering process. 

\begin{figure}[t]
\centering

\begin{minipage}{0.23\linewidth}
\centering
\begin{tikzpicture}[scale=0.9]
  \coordinate (i1) at (-1, 1);
  \coordinate (i2) at (-1,-1);
  \coordinate (f1) at ( 1, 1);
  \coordinate (f2) at ( 1,-1);
  \coordinate (v)  at ( 0, 0);

  \draw (i1) -- (v) -- (f2);
  \draw (i2) -- (v) -- (f1);

  \node[left]  at (i1) {$m_1$};
  \node[left]  at (i2) {$m_2$};
  \node[right] at (f1) {$m_1$};
  \node[right] at (f2) {$m_2$};
\end{tikzpicture}
\caption*{(a)}
\end{minipage}\hfill
%
\begin{minipage}{0.23\linewidth}
\centering
\begin{tikzpicture}[scale=0.9]
  \coordinate (i1) at (-1, 1);
  \coordinate (i2) at (-1,-1);
  \coordinate (f1) at ( 1, 1);
  \coordinate (f2) at ( 1,-1);
  \coordinate (vL) at (-0.4,0);
  \coordinate (vR) at ( 0.4,0);

  \draw (i1) -- (vL) -- (i2);
  \draw (vR) -- (f1);
  \draw (vR) -- (f2);
  \draw (vL) -- (vR);

  \node[above] at (0,0) {$m_{i,s}$};
  \node[left]  at (i1) {$m_1$};
  \node[left]  at (i2) {$m_2$};
  \node[right] at (f1) {$m_1$};
  \node[right] at (f2) {$m_2$};
\end{tikzpicture}
\caption*{(b)}
\end{minipage}\hfill
%
\begin{minipage}{0.23\linewidth}
\centering
\begin{tikzpicture}[scale=0.9]
  \coordinate (TL) at (-1,  1);
  \coordinate (TR) at ( 1,  1);
  \coordinate (BL) at (-1, -1);
  \coordinate (BR) at ( 1, -1);
  \coordinate (vT) at ( 0,  1);
  \coordinate (vB) at ( 0, -1);

  \draw (TL) -- (vT) -- (TR);
  \draw (BL) -- (vB) -- (BR);
  \draw (vT) -- (vB);

  \node[right] at (0,0) {$m_{j,t}$};

  \node[above] at (TL) {$m_1$};
  \node[above] at (TR) {$m_1$};
  \node[below] at (BL) {$m_2$};
  \node[below] at (BR) {$m_2$};
\end{tikzpicture}
\caption*{(c)}
\end{minipage}\hfill
%
\begin{minipage}{0.23\linewidth}
\centering
\begin{tikzpicture}[scale=0.9]
  \coordinate (i1L) at (-1,  0.7);
  \coordinate (i2L) at (-1, -0.7);
  \coordinate (vT)  at ( 0,  0.7);
  \coordinate (vB)  at ( 0, -0.7);
  \coordinate (f1R) at ( 1,  1.0);
  \coordinate (f2R) at ( 1, -1.0);

  \draw (i1L) -- (vT);
  \draw (i2L) -- (vB);

  \draw (vT) -- (vB);

  \draw (vT) -- (f2R);
  \draw (vB) -- (f1R);

  \node[left]  at (i1L) {$m_1$};
  \node[left]  at (i2L) {$m_2$};
  \node[right] at (f1R) {$m_1$};
  \node[right] at (f2R) {$m_2$};
  \node[left]  at (0,0) {$m_{k,u}$};
\end{tikzpicture}
\caption*{(d)}
\end{minipage}

\caption{Scattering topologies contributing to $\mathcal M$.}
\label{fig:DiagramTopologies}
\end{figure}

\begin{figure}[h!]
\centering
\begin{subfigure}[b]{0.48\textwidth}
    \includegraphics[width=\textwidth]{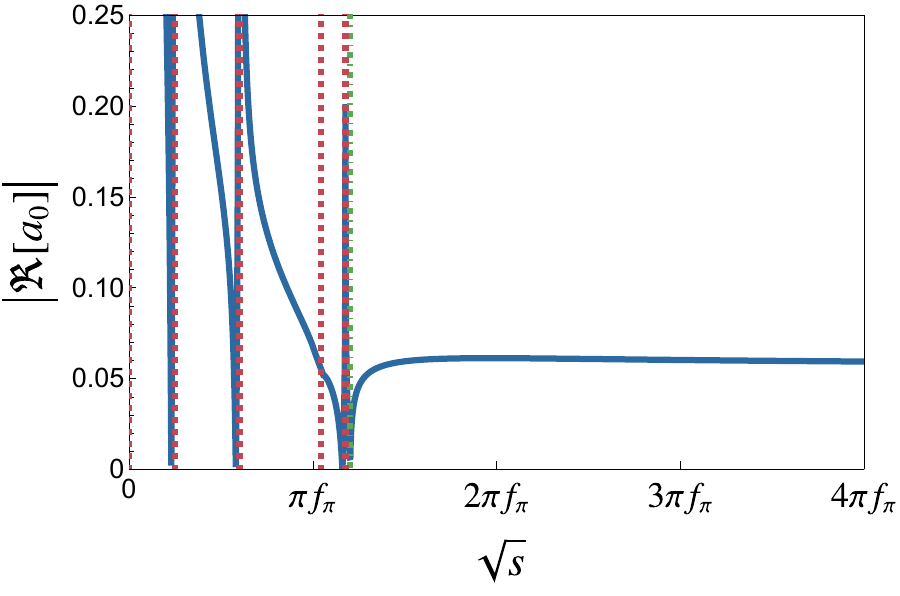}
    \caption{$XX\rightarrow XX$}
    \label{fig:UnitarityXXXX}
\end{subfigure}
\hfill
\begin{subfigure}[b]{0.48\textwidth}
    \includegraphics[width=\textwidth]{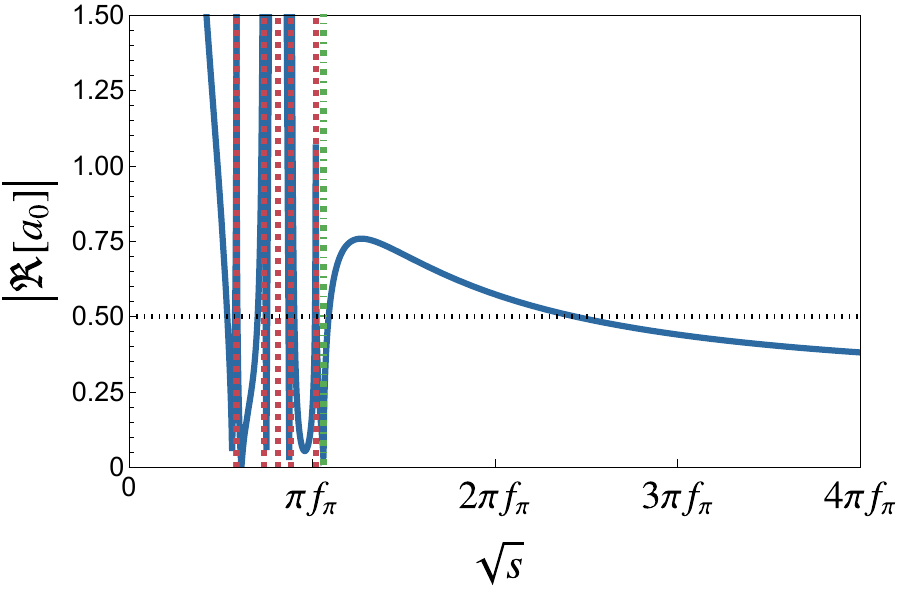}
    \caption{$\eta'\sigma\rightarrow \eta'\sigma$}
    \label{fig:Unitarityeta2sigma2}
\end{subfigure}
\caption{The $|\Re(a_0)|$ vs $\sqrt{s}$. For $XX\rightarrow XX$ scattering in \cref{fig:UnitarityXXXX} and $\eta'\sigma\rightarrow \eta'\sigma$ in \cref{fig:Unitarityeta2sigma2} for $F=4$. The green dot-dashed line represents the minimum kinematically accessible centre of mass energy, and the dashed red lines represents (kinematically inaccessible) poles in the 4-point amplitude. Unitarity bounds are set by scanning from the minimum kinematically accessible energy to the EFT cutoff at $\Lambda_\text{cutoff}=4\pi f_\pi$. If the maximum exceeds $|\Re(a_0)|>0.5$, the point is non-unitary which is indicated by a horizontal dotted line in \cref{fig:Unitarityeta2sigma2}. Any poles in the kinematically accessible region are excluded from the scan by \cref{eq:PoleCutout}, which do not appear for the cases shown in this figure. Since the $|\Re(a_0)|$ does exceed 0.5 in the kinematically accessible region, this point does not obey tree-level perturbative unitarity. This benchmark is shown for $m_\sigma^2=1\times10^6\;\text{(GeV)}^2,\; m_{\eta'}^2=1.06\times10^7\;\text{(GeV)}^2,\; m_X^2=5.8\times10^6\;\text{(GeV)}^2$ and $f_\pi=1270\;\text{GeV}$. Equivalently in terms of Lagrangian parameters, $m_\Phi^2=5\times10^5\;(\text{GeV})^2,\,c=6.54,\lambda_\sigma=3.89$ and $\lambda_a=0.309$.
}
\label{fig:pw-unit}
\end{figure}

Across the kinematically accessible regime, perturbative unitarity requires
\begin{align} 
    |\Re(a_0)|\leq\frac{1}{2} \,,
\end{align}
where the $a_0$, the zeroth partial wave, is a function of the center of mass energy, $a_0=a_0(s)$. We require this condition holds for all $s$ in the regime that satisfies
\begin{align}
    \left(m_1^2 + 2\sqrt{m_1^2 m_2^2} + m_2^2\right)^{1/2} \leq s\leq \Lambda_\text{cutoff}=4\pi f_\pi,
\end{align}
where $\Lambda_\text{cutoff}$ is the cutoff of the EFT. We depict the behavior of the zeroth partial wave across the kinematically relevant region in Fig.~\ref{fig:pw-unit}.

Our expressions for the tree-level $2\to2$ meson scattering amplitudes may naively contain poles. Proper treatment of these divergences would require computing higher-loop corrections to the elastic scattering. To ensure that our tree-level approximation is not falsely flagging loss of unitarity due to the inclusion of an unphysical pole contribution, we follow the procedure in \cite{Goodsell:2018tti}. That is, we cut out the entire range in which
\begin{align}\label{eq:PoleCutout}
    \left|1-\frac{s}{m_{i,s}^2}\right|\geq 0.25 \,,
\end{align}
where $m_{i,s}$ runs over all possible s-channel propagators (in the kinematically accessible region) denoted by $i$.

In addition, there may also be t- and u-channel poles appearing in the amplitude. 
We expect the t-poles to appear when
\begin{align}
   \! m_{j,t}^2
        &= 2m_1^2+2m_2^2-s_\text{t-pole}
            -\frac{m_1^2+m_2^2-2m_1^2m_2^2}{s_\text{t-pole}} \,.
\end{align}
Rearranging gives
\begin{align}
    s_\text{t-pole}
        &=m_1^2+m_2^2
            - \dfrac{ m_{j, t}^2
                \pm \sqrt{4m_1^2-m_{j,t}^2}\sqrt{4m_2^2-m_{j,t}^2} }2\,. 
\end{align}
The u-poles appear when
\begin{align}
     m_{k,u}^2
        &= \frac{(m_1^2-m_2^2)^2}{s_\text{u-pole}}
    & \Rightarrow s_\text{u-pole}
        &=\frac{(m_1^2-m_2^2)^2}{m_{k,u}^2}\,,
    \\
    m_{k,u}^2
        & = 2m_1^2+2m_2^2
            -s_\text{u-pole}
     &\Rightarrow s_\text{u-pole}
        &=2(m_1^2+m_2^2)-m_{k,u}^2 \,.
\end{align}
We apply the same treatment as \cref{eq:PoleCutout} for these poles if they are in the kinematically accessible region.
Additionally, the total cross-section is badly defined when: 
\begin{equation}
    m_1^2 = m_2^2 \quad \text{if} \quad m_{k,u}^2 = 0 \,.
\end{equation}

In which case, we ignore that particular scattering process and only consider bounds from the other processes. We note that while our estimation of the unitarity bound is approximate, the restriction on parameter space is ultimately unavoidable. In contrast, our perturbativity criteria could in principle be ameliorated with e.g. going to higher order in the resummation scheme.

\subsection{Vacuum Stability}
In the case where $F>4$ and therefore the dimension of the determinant term exceeds 4, requiring the $\eta'$ mass to be (strictly) positive at $\sigma=f_\pi$ sets the sign of $c$ in Eq.~(\ref{eq:lsm}) and guarantees the potential will turn downwards for large values of $\sigma$.  This causes a boundedness from below issue if the potential drops below the minimum at $\sigma=f_\pi$ below the cutoff $\Lambda=4\pi f_\pi$. We numerically check that this does not occur in the regime of validity of the EFT using an ``effective boundedness from below'' condition:
\begin{align}
   & V_0^\text{LSM}(\sigma=f_\pi)
        <V_0^\text{LSM}(\sigma=4\pi f_\pi) 
    \label{eq:bdd} \,.
\end{align}
Note that if the potential is in danger of losing boundedness from below, $V_0^{\rm LSM}(\sigma =4\pi f_\pi)$ will be the lower than any $V_0^{\rm LSM}(\sigma < 4\pi f_\pi)$, and so it is sufficient to evaluate and check the value of $V_0^{\rm LSM}$ only at the cutoff $4\pi f_\pi$. 
We are not concerned about the EFT naively indicating that $V_0^{\rm LSM}$ develops a deeper mininum for field values  above $\Lambda = 4\pi f_\pi$. In this case, UV physics would enter and could upturn the potential, to which we remain agnostic.

We also explicitly require an algebraic vacuum stability condition from~\cite{Hansen:2017pwe} and used in~\cite{Pasechnik:2023hwv},
    \begin{align}
    \lambda_a > - \frac1{ F -1 } \lambda_\sigma \,,
    \end{align}
in addition to our numerical requirement for effective boundedness from below in Eq.~(\ref{eq:bdd}).

\section{The Phase Transition}\label{sec:PT}

Here we lay out the ingredients of our analysis that translate the effective potential into an observable GW signal. We follow standard methods in the literature~\cite{Randall:2006py, Gelmini:1988sf}. The GW signal originates from the dynamics of vacuum bubbles colliding and interacting with the surrounding plasma during a cosmological FOPT. The surface tension of these bubbles reflects the field configuration that interpolates across the barrier between the false and true minima. 

The quantity that governs the nucleation temperature and thermal parameters entering the GW spectrum is the Euclidean bounce solution, obtained from the finite-$T$ action. The Euclidean action is given by:
    \begin{align}
    S_3 (T)
        &= 4\pi \displaystyle\int_0^\infty dr\,r^2 
            \left[ \frac12 \left( \frac{ d\sigma}{dr} \right)^2 
                + V_{\rm eff}( \sigma, T) \right] \,.
    \end{align}
The corresponding equations of motion and boundary conditions are
    \begin{align}
    \frac{ d^2 \sigma }{ dr^2 }
        + \frac2r &\frac{ d\sigma }{ dr }
        = \frac{ \partial V_{\rm eff}}{\partial \sigma}\,,
    & \left. \frac{ d\sigma }{dr } \right|_{r=0}
        =0\,,
    &\qquad \lim_{r\to \infty} \sigma = 0 \,.
    \end{align}
This equation of motion was solved using the \texttt{CosmoTransitions} software package~\cite{Wainwright:2011kj}. For the subsequent numerical analysis solving the coupled evolution equations for the thermally dressed masses, evolving the effective potential, locating the minima, and extracting the phase transition parameters we use a combination of \texttt{CosmoTransitions} and dedicated Python/Mathematica routines. This code is publicly available for the community at \href{https://github.com/miarobin/Eta-Prime-GWs}{\faGithub}.

To determine the temperature at which the phase transition occurs, we require the tunnelling rate per unit volume, 
    \begin{align}
    \Gamma
        &= T^4 \left( \frac{ S_3(T) }{ 2\pi T } \right)^{3/2} e^{-  S_3 (T)/T} \,. 
    \label{eq:tun-rate}
    \end{align}
To obtain the temperature of the phase transition, meaning the temperature at which to evaluate the thermal parameters, we follow the procedure in~\cite{Pasechnik:2023hwv}. This favors the percolation temperature $T_p$ over the typical nucleation temperature $T_N$. Whereas $T_N$ results from approximately equating the nucleation rate $\Gamma$ to the Hubble volume, the percolation temperature $T_p$ is instead defined as the temperature when the probability of being in the false vacuum reaches a percolation threshold~\cite{Guth:1979bh, Guth:1981uk, Ellis:2018mja},  
    \begin{align}
    P(T_p) = e^{-I(T_p)} \sim 0.7 \,,
    \end{align}
where $I(T)$ is a weight function
    \begin{align}
    I(T)
        &= \! \frac{ 4\pi }3 \displaystyle\int_T^{T_c} \!\! dT' 
                \frac{ \Gamma(T') }{ H(T') {T'}^4} 
                \left( \displaystyle\int_T^{T'} \!\!\! dT''
                    \frac{ v_w(T'') }{ H(T'')} \right)^3 ,
    \end{align}
and $\Gamma(T')$ is given in Eq.~\eqref{eq:tun-rate}. Here, $v_w$ is the bubble wall velocity. 
During a domination radiation era assuming a roughly constant wall velocity $ v_{w}\sim1$, the weight function is given by:
\begin{align}
    I_{R}(T) =  12 \pi (M_{pl} \xi_{g})^4 \int_{T}^{T_c} \frac{d T' \Gamma(T')}{T^{'6}} \left( \dfrac{1}{T} - \dfrac{1}{T'}\right)^3
\end{align}
Where \( M_{pl}\) is the Planck mass, and \(\xi_{g} = \sqrt{30/(\pi^2 g_{*})}\).
The thermal parameters $\alpha$ and $\beta$ follow directly from the bounce action. The strength parameter is 
    \begin{align}
     \alpha
        &= \frac1{ \rho_{\rm rad} (T_p) }\left[\Delta V(T_p) -  T_p \left.\frac{ d \Delta V }{dT } \right|_{T=T_p} \right] \,,
    \label{eq:alpha}
    \end{align}
where $\Delta V(T)$ is the potential difference between the false and true minima. The radiation density is $\rho_{\rm rad}(T_p)=g_* \pi^2 T_p^4 /30$, with  $g_*$ the effective number of relativistic degrees of freedom at $T_p$. We set $g_*(T_p)=g_{*,\text{SM}}(T_p)+2(N^2-1)+\frac{7}{2}NF$, where $g_{*,\text{SM}}= 106.75$ for the $\sim$TeV scale phase transitions considered here. 
The inverse duration parameter is
    \begin{align}
    \frac\beta{ H_p}
        &= T_p \left. \frac{ d (S_3 / T )}{ dT } \right|_{T=T_p} \,,
    \end{align}
where $H_p \equiv H(T_p)$ is Hubble evaluated at the percolation temperature. 

Finally, to compute the GW signal we also require the bubble wall velocity. We employ two methods to calculate this. The first uses the Local Thermal Equilibrium (LTE) approximation which is an estimate of the maximum speed of the bubble wall. This method is detailed in \cite{Ai:2023see}. Note that a lower bound on the wall velocity could be also be computed with the ballistic limit \cite{Ai:2024btx}. The second method, which we denote as the Large Change in Degrees of Freedom (LCDF) method, makes use of the limit in which the number of degrees of freedom in the plasma drops significantly through the phase transition. We expect the LCDF method to be valid when
\begin{align}
    2(N^2-1)+\frac{7}{2}NF\gg2F^2 \,.
    \label{eq:dof-conf}
\end{align}
This results from comparing the active degrees of freedom above confinement on the LHS (from $N^2-1$ gluons, $NF$ quarks) with the active degrees of freedom below confinement on the RHS (from $2F^2$ mesons). This second method is detailed in \cite{Sanchez-Garitaonandia:2023zqz}. However, introducing the SM degrees of freedom in the plasma dilutes the relative drop across the bubble wall. As a result, the limits taken are less likely to be applicable for the cases discussed here. We still include it for interest and note that this method is most valid for $F\lesssim N$. 
The signal resulting from employing both choices of $v_w$ will be detailed and compared in~\cref{sec:results}. To evaluate the wall velocities, we use the code snippet provided in \cite{Ai:2023see} which is slightly modified to the LCDF method. This is also included in the GitHub repository.

The total GW spectrum is the sum of the sound wave contribution $\Omega_{\rm sw}$, the turbulence contribution $\Omega_{\rm turb} $, and the subdominant contribution from bubble wall collisions $\Omega_{\rm col}$. We neglect the latter as it is expected to contribute at the percent level. 
The sound wave contribution is: 
    \begin{align}
    h^2 \Omega_{\rm sw} (f)
        &= 2.59 \times 10^{-6} h^2 
            \left( \frac{ 100 }{ g_* } \right)^{1/3} 
             \left( \frac{  \kappa_{\rm sw} \alpha} { 1 + \alpha  }\right)^2
            \left( \frac{ H_p }\beta \right)
            v_w 
            S_{\rm sw} (f)\,,
        \label{eq:omega-sw}
    \end{align}
where 
$\kappa_{\rm sw}$ is the efficiency of converting vacuum energy into bulk field motion. The spectral shape is
    \begin{align}
    S_{\rm sw} (f)
        &= \left( \frac f { f_{\rm sw }} \right)^3
            \left( \frac7{ 4 + 3 (f/f_{\rm sw})^2 } \right)^{7/2}\,,
    \end{align}
where $f_{\rm sw}$ is the peak frequency observed today, 
    \begin{align}
    f_{\rm sw}
        &= (8.876 \text{ }\mu\text{Hz} ) 
            \left( \frac{ g_*} {100} \right)^{1/6}
            \left( \frac{ T_p }{ 100 \text{ GeV } } \right)
            \left( \frac1{ v_w } \right)
            \left( \frac\beta H_p \right) 
        \,.
    \end{align}
The turbulence contribution is
    \begin{align}
    h^2 \Omega_{\rm turb} (f)
        &= 3.35 \times 10^{-4}
            \left( \frac {H_p}\beta \right) 
            \left( \frac{ \kappa_{\rm turb} \alpha}{1 + \alpha} \right)^{3/2}
            \left( \frac{100}{ g_*} \right)^{1/3}
            v_w \,\, S_{\rm turb}(f)\,,
        \label{eq:omega-turb}
    \end{align}
where $\kappa_{\rm turb}$ is the efficiency factor for turbulence, analogous to $\kappa_{\rm sw}$ for sound waves. We use $\kappa_{\rm turb} = 0.05 \kappa_{\rm sw}$, following~\cite{Caprini:2015zlo}. The spectral shape is now
    \begin{align}
    S_{\rm turb}(f)
        &= \frac{ (f/f_{\rm turb})^3 }
               { \left[ 1 + f/f_{\rm turb}\right]^{11/3} 
                \left( 1 + 8\pi f / h_p \right)} \,,
    \end{align}
where $h_p$ is the Hubble frequency redshifted from $T_p$, and $f_{\rm turb}$ is the peak frequency for the turbulence signal, given by:
    \begin{align}
    f_{\rm turb}
        &= 27 \mu\text{Hz} \frac1{ v_w } 
            \left( \frac\beta {H_p } \right) 
            \left( \frac{ T_p }{ 100 \text{ GeV} } \right)
            \left( \frac{ g_* }{ 100 } \right)^{1/6}  
    \\
    h_p
        &= 16.6\,\mu\text{Hz} \, \left( \frac{ T_p }{ 100 \text{ GeV} } \right) \left( \frac{ g_* }{ 100 }\right)^{1/6} \,.
    \end{align}
With this we can present the full GW power spectrum:
    \begin{align}
    h^2 \Omega(f)
        &= \underbrace{h^2 \Omega_{\rm sw}(f)}_{\text{Eq.~}\eqref{eq:omega-sw}}
            + \underbrace{h^2 \Omega_{\rm turb}(f)}_{\text{Eq.~}\eqref{eq:omega-turb}}
            + \underbrace{h^2 \Omega_{\text{col}}(f)}_{\text{ignored}} \,.
    \end{align}
This is the GW spectrum $h^2 \Omega (f)$ presented in our results and used to assess observability of the signal.

\section{Results}
\label{sec:results}

In this section we present the numerical results, emphasizing the main features of the phase transition and its corresponding GW signatures across the region of parameter space that satisfies our EFT consistency criteria. We also evaluate the detection prospects of these signals in proposed future GW observatories including the Laser Interferometer Space Antenna (LISA)~\cite{LISA:2017pwj, Baker:2019nia}, the Big Bang Observer (BBO)~\cite{Crowder:2005nr, Corbin:2005ny, Yagi:2011wg, Harry:2006fi}, the Atomic Experiment for Dark Matter and Gravity Exploration in Space (AEDGE)~\cite{AEDGE:2019nxb}, the Deci-hertz Interferometer Gravitational wave Observatory (DECIGO)~\cite{Seto:2001qf}, and the Einstein Telescope (ET)~\cite{Punturo:2010zz}. 

Our GW signal was obtained by scanning across the values of the tree-level mass parameters in the ranges:
\begin{align}
    m_\sigma^2&\in[1,10]\times10^6\;\left(\text{GeV}\right)^2\\
    m^2_X&\in[1,25]\times 10^6 \;\left(\text{GeV}\right)^2\\
    f_\pi&\in[0.5,1.5]\sqrt{\frac{F}{2}
    }\times 10^3\;\text{GeV}
\end{align}
and
\begin{align}
    \text{if}\;\; &F\leq 4:m_{\eta'}^2\in[1,25]\times10^6\;\left(\text{GeV}\right)^2 \\
    \text{if}\;\; &F>4: m_{\eta'}^2\in[1,25]\frac{3F^3}{8(4\pi)^{F-4}}\times 10^4 \;\left(\text{GeV}\right)^2 
\end{align}

\begin{figure}[h!]
\centering
\includegraphics[width=.85\textwidth, clip=true, trim=.6cm 0cm 0cm 0cm]{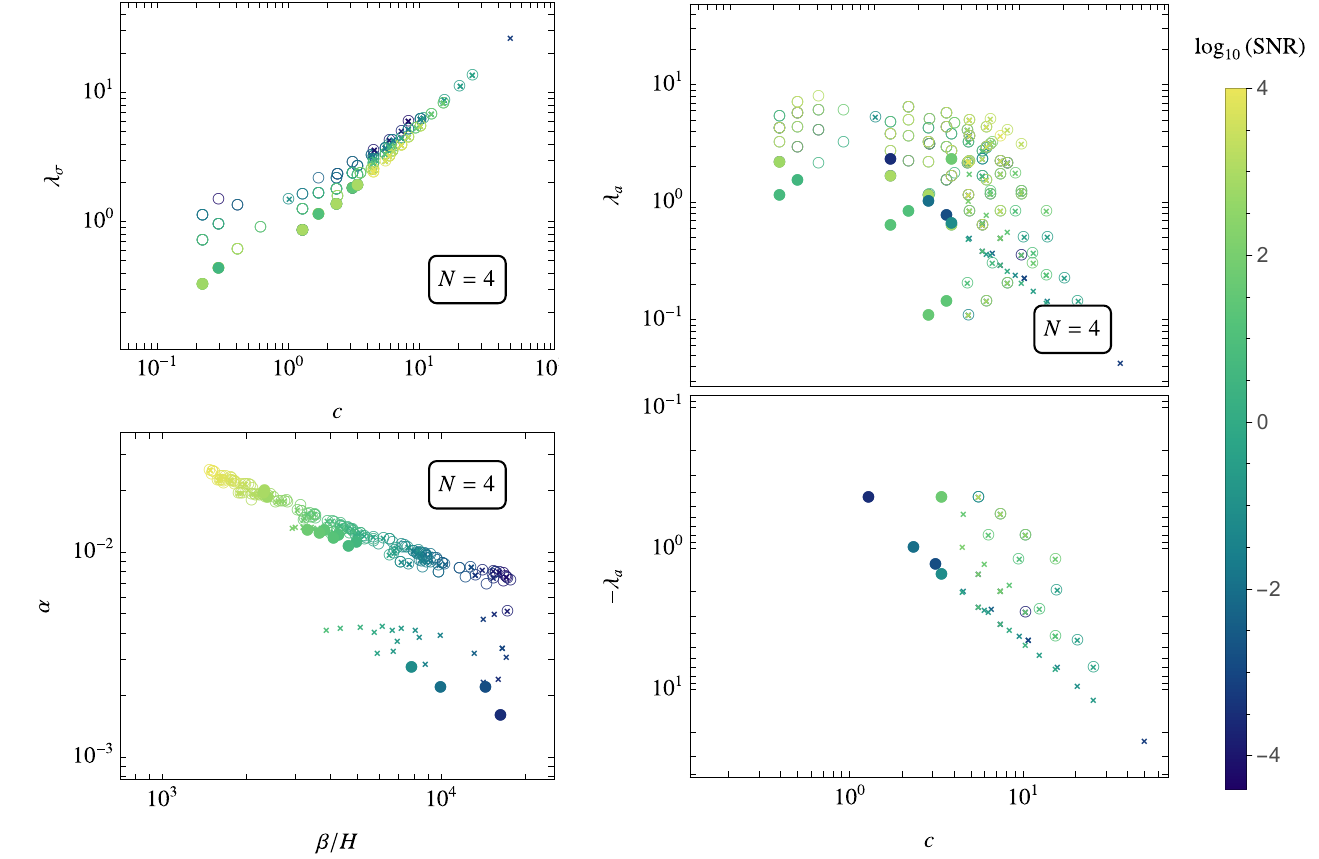}

\vspace{.2cm}
\includegraphics[width=.85
\textwidth, clip=true, trim=.6cm 0cm 0cm 0cm]{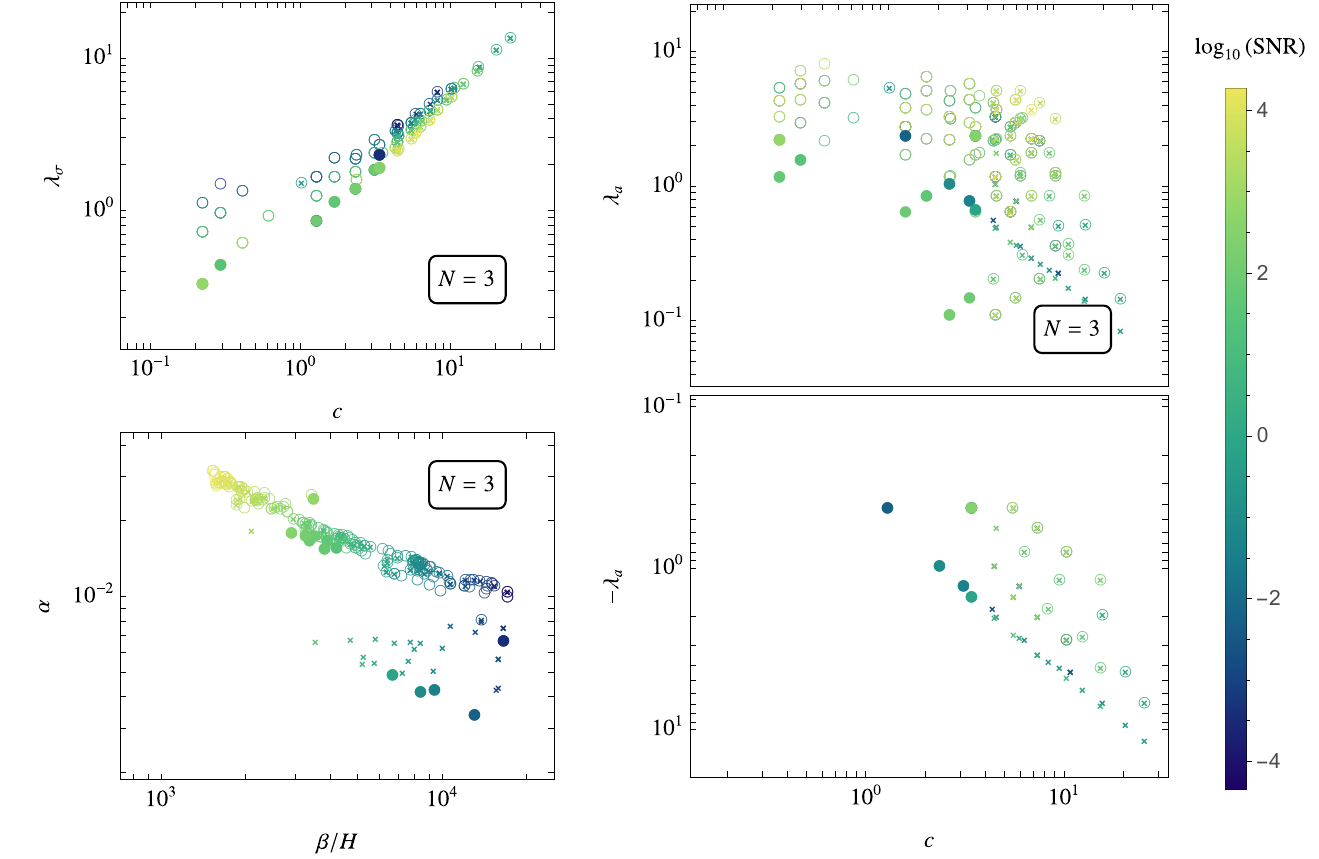}
\vspace{-.25cm}
\caption{The SNR for observing the GW signal in the BBO detector for different couplings in the potential that yield FOPTs. In all panels, $F=4$, $\xi=2$, and $v_w$ is calculated using the LTE method. The top set of plots shows $N=4$, while the bottom set shows $N=3$. The color scaling indicates SNR for each parameter point. In the top left panel for each set, we present $\lambda_\sigma$ vs. $c$. In the right column for each set, we present $\pm\lambda_a$ vs. $c$ (separated to show log scaling). In the bottom left panel, we plot the thermal parameters $\alpha$ vs. $\beta/H$. 
Filled points on the plot pass all EFT consistency requirements. Open circles are points that fail the perturbativity requirement and crosses represent points that fail the unitarity requirement. When a point fails both, the circle and cross are overlaid. 
}
\label{fig:snr-f4}
\end{figure}

\begin{figure}[h!]
\centering
\includegraphics[width=.85\textwidth, clip=true, trim=.6cm 0cm 0cm 0cm]{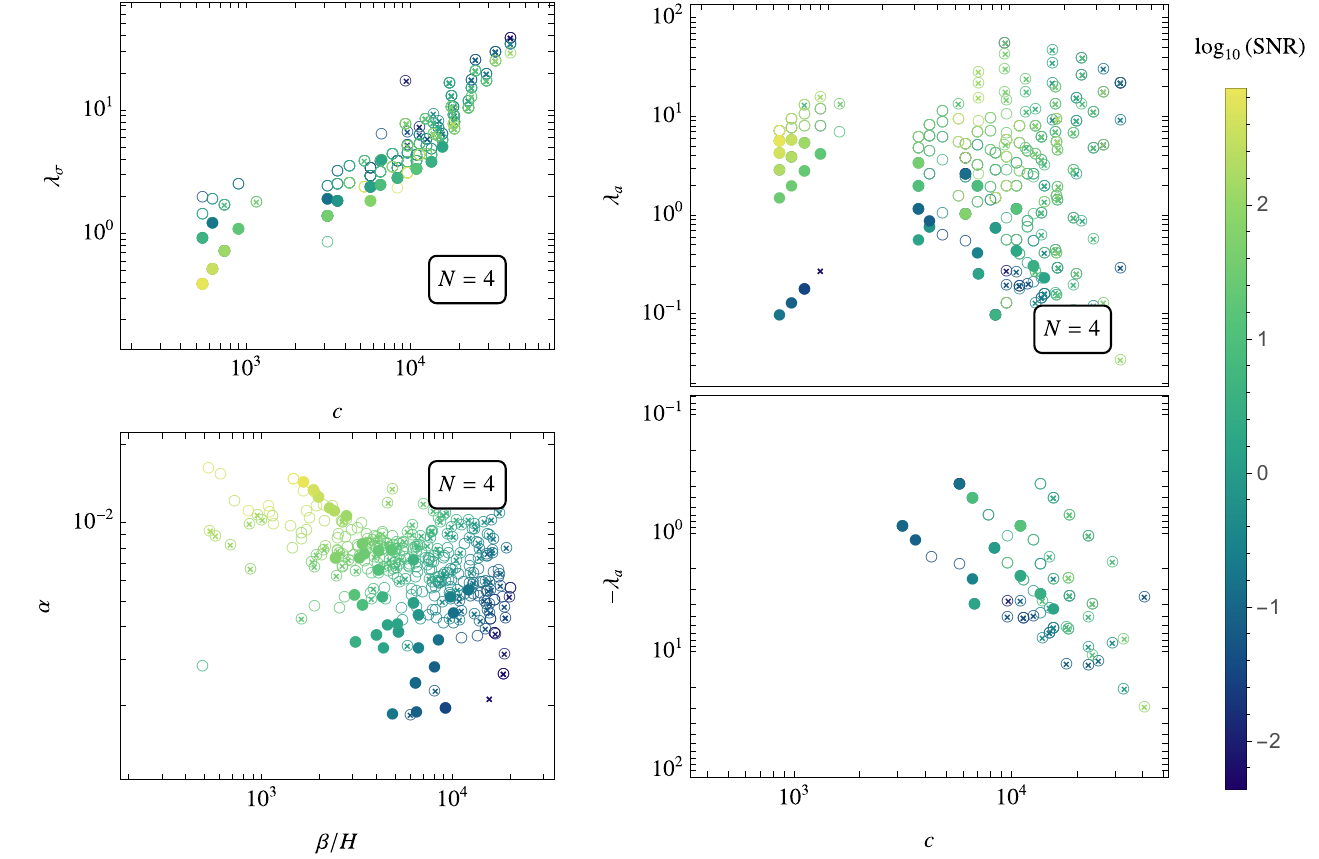}

\vspace{.2cm}
\includegraphics[width=.85\textwidth, clip=true, trim=.6cm 0cm 0cm 0cm]{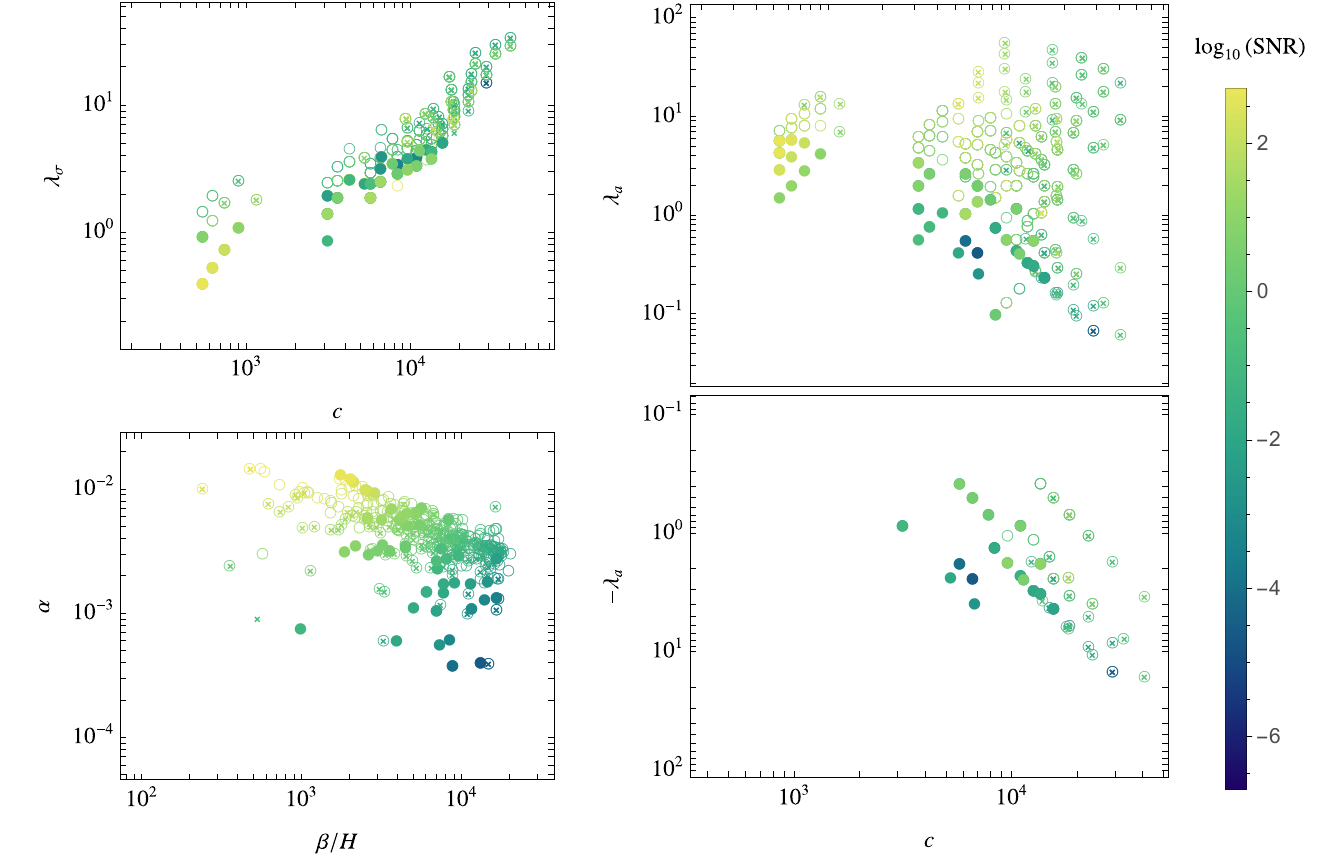}
\vspace{-.25cm}
\caption{The SNR for observing the GW signal in the BBO detector for different couplings in the potential that yield FOPTs. In all panels, $F=3$, $\xi=1$, and $v_w=1$. The top set of plots are for $N=4$, while the bottom set are the results when the Polyakov loop contributions are turned off. The coupling $c$ is in units of $\text{GeV}$ when $F=3$. Like in Fig.~\ref{fig:snr-f4}, filled points on the plot pass all EFT consistency requirements; open circles are points that fail the perturbativity requirement and crosses represent points that fail the unitarity requirement; and when a point fails both, the circle and cross are overlaid. 
}
\label{fig:snr-f3}
\end{figure}

In Figs.~\ref{fig:snr-f4} and~\ref{fig:snr-f3}, we present the signal-to-noise ratio (SNR) for the gravitational wave signal relative to the detector sensitivity curve of BBO~\cite{Crowder:2005nr, Corbin:2005ny, Harry:2006fi, Yagi:2011wg, Thrane:2013oya}. The SNR is given by~\cite{Allen:1997ad,Maggiore:1999vm}
\begin{align}
\text{SNR}
    &= \sqrt{ \frac {T_{\rm obs}}{ \rm sec. } 
            \displaystyle\int df \left( \frac{ h^2 \Omega_{\text{signal}}(f)}{ h^2 \Omega_{\rm sens}(f)} \right)} \,,
    \label{eq:snr}
\end{align}
where $T_{\rm obs}$ is the observation time in seconds, $h^2\Omega_{\rm signal}(f)$ is the predicted GW power spectrum, $h^2 \Omega_{\rm sens}(f)$ is the GW detector sensitivity curve, and the integration region spans the frequencies over which the detector is sensitive. 
Fig.~\ref{fig:snr-f4} shows the SNR obtained for BBO across slices of EFT parameter space that yield FOPTs. For this SNR, we assumed an observation time period of $T_{\rm obs} = 3$ years. Rather than scanning directly over the LSM couplings, we scan over the physical meson masses at zero temperature. This implicitly determines the LSM couplings $\lambda_\sigma, \lambda_a, m_\Phi^2,$ and $c$. In Fig.~\ref{fig:snr-f4}, we fix $F=4$, set $\xi=2$,\footnote{Recall $\xi$ sets the confinement temperature relative to $f_\pi$, defined in Eq.~\eqref{eq:xi}.} and use the bubble wall velocity $v_w$ obtained using the LTE prescription. In all plots, the color scale encodes the resulting SNR in BBO. Different markers distinguish parameter points that satisfy the EFT consistency conditions from those that fail perturbativity, unitarity, or both. The upper set of panels corresponds to $N=4$, while the lower set shows the analogous scan for $N=3$. In each case, the top left plot displays the $\lambda_\sigma$-$c$ plane, the right column the $\pm\lambda_a$-$c$ plane, and the bottom-left plot shows the thermal parameters $\alpha$ and $\beta/H$. This highlights which regions of coupling space both support a FOPT with a potentially detectable GW signal while remaining theoretically under control. 

Fig.~\ref{fig:snr-f3} shows the corresponding SNR results for $F=3$, now with $\xi=1$ and a fixed bubble-wall velocity $v_w=1$, again using an observation time period of $T_{\rm obs} = 3$ years. As before, we scan over physical meson masses at zero temperature, and the corresponding LSM couplings follow from those inputs. The upper panels show the case with the Polyakov-loop effects for $N=4$, while the lower panels show the same scan with the Polyakov-loop sector ($V_{\rm PLM}$ and $V_{\rm med}$) switched off. This comparison isolates the impact of gluonic effects on the viable parameter space. Together, Fig.~\ref{fig:snr-f4} and~\ref{fig:snr-f3} illustrate how the SNR varies across the EFT parameter space once the consistency constraints are imposed, and how turning off or modifying the Polyakov-loop sector affects the surviving regions and their associated GW signals.

\begin{figure}[h!]
    \centering
    \begin{subfigure}[b]{0.48\textwidth}
        \centering
        \includegraphics[width=\textwidth, clip=true, trim=.0cm .0cm .0cm .6cm]{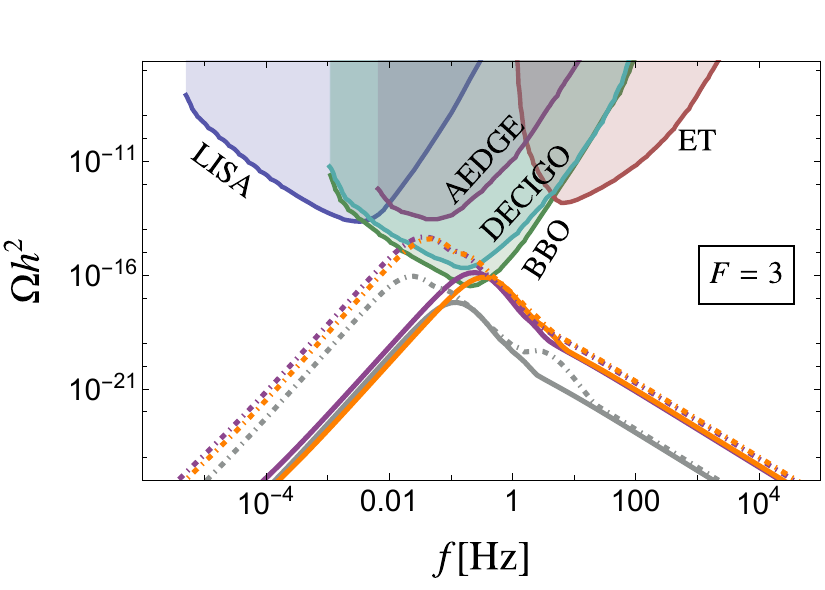} 
        \label{fig:PassFailvwN3F3}
    \end{subfigure}
    \begin{subfigure}[b]{0.48\textwidth}
        \centering
        \includegraphics[width=\textwidth, clip=true, trim=.cm .0cm .0cm .6cm]{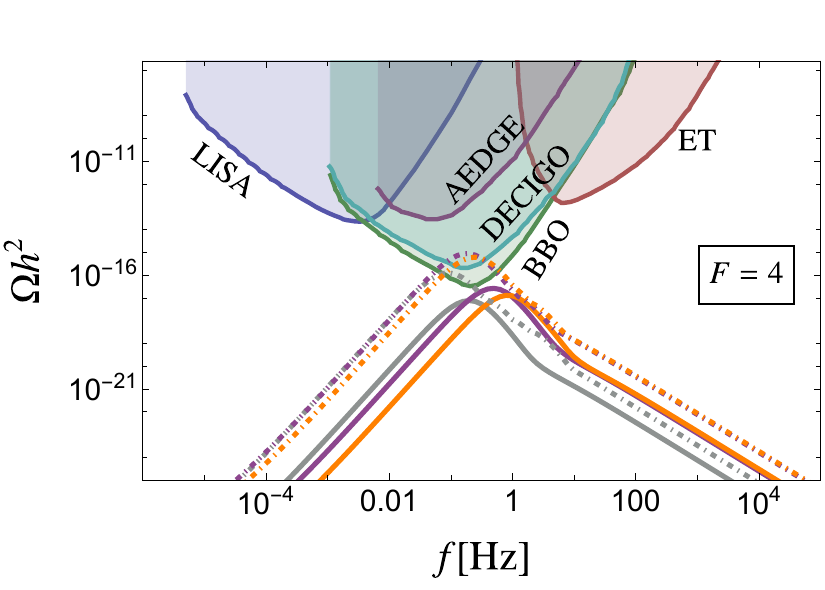} 
        \label{fig:PassFailvwN3F4}
    \end{subfigure}
    \\
    \vspace{-1.5em}
    \begin{subfigure}[b]{0.48\textwidth}
        \centering
        \includegraphics[width=\textwidth, clip=true, trim=.0cm .0cm .0cm .6cm]{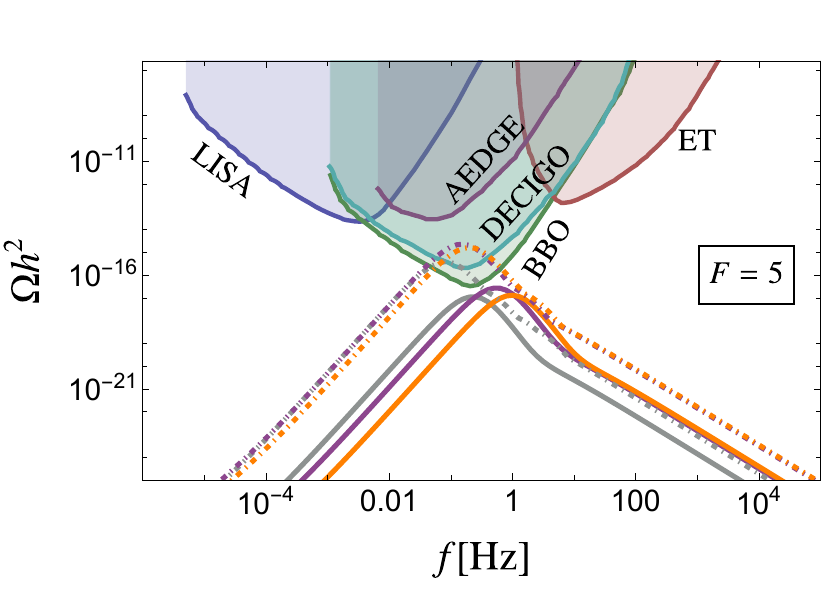} 
        \label{fig:PassFailvwN3F5}
    \end{subfigure}
    \begin{subfigure}[b]{0.48\textwidth}
        \centering
        \includegraphics[width=\textwidth, clip=true, trim=.cm .0cm .0cm .6cm]{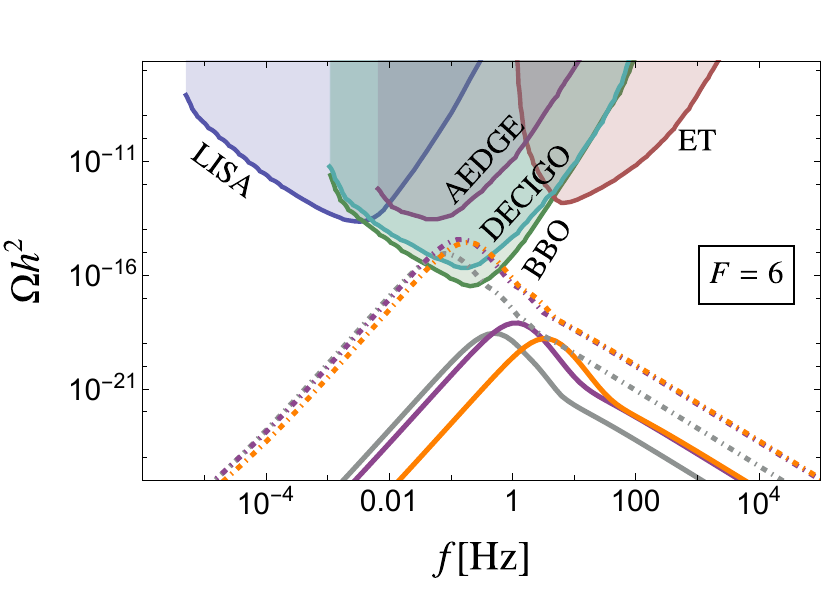} 
        \label{fig:PassFailvwN3F6}
    \end{subfigure}
    \vspace{-1.5em}
    \caption{Comparison of signal strength against $v_w$ for $N=3$ and $\xi=1$. Grey indicates $v_w=1$, i.e. runaway, purple LTE \cite{Ai:2023see} and orange LCDF \cite{Sanchez-Garitaonandia:2023zqz}. The dot-dashed line includes all tested points. The solid line only includes those points for which the potential is unitary, and our perturbativity criteria hold.}
    \label{fig:PassFailVwallN3}
\end{figure}

\begin{figure}[h!]
    \centering
    \begin{subfigure}[b]{0.45\textwidth}
        \centering
        \includegraphics[width=\textwidth, clip=true, trim=.0cm .0cm .0cm .6cm]{Figures/N4F3Envelope_xi1_vwScan.pdf} 
        \label{fig:PassFailvwN4F3}
    \end{subfigure}
    \begin{subfigure}[b]{0.45\textwidth}
        \centering
        \includegraphics[width=\textwidth, clip=true, trim=.cm .0cm .0cm .6cm]{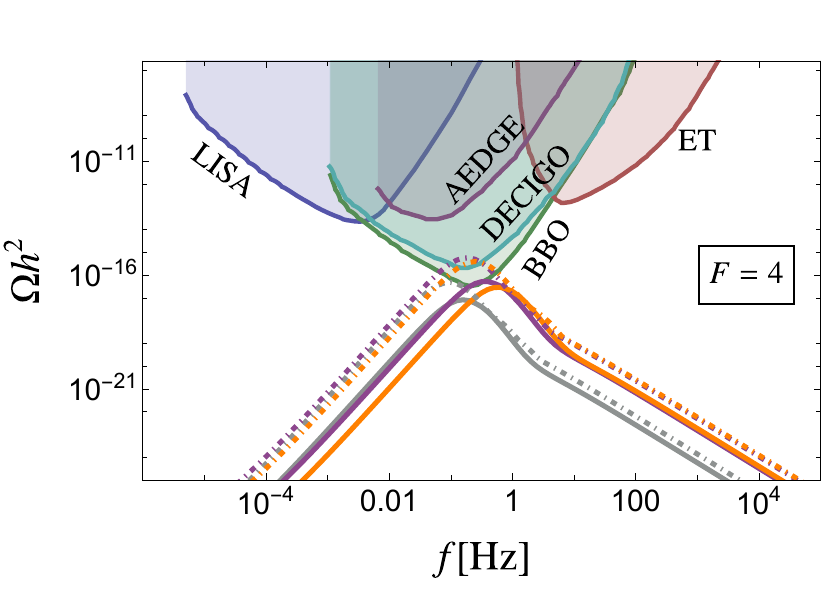} 
        \label{fig:PassFailvwN4F4}
    \end{subfigure}
    \\
    \vspace{-1.5em}
    \begin{subfigure}[b]{0.45\textwidth}
        \centering
        \includegraphics[width=\textwidth, clip=true, trim=.0cm .0cm .0cm .6cm]{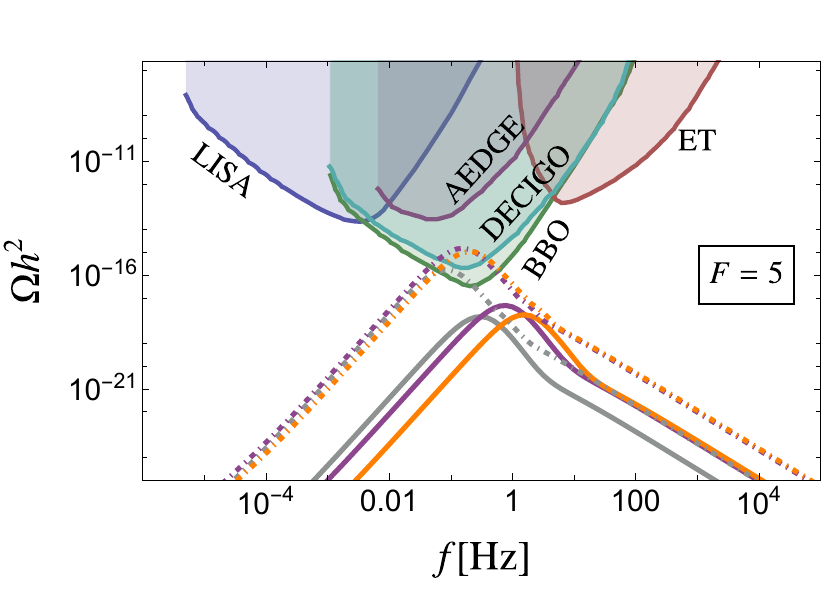} 
        \label{fig:PassFailvwN4F5}
    \end{subfigure}
    \begin{subfigure}[b]{0.45\textwidth}
        \centering
        \includegraphics[width=\textwidth, clip=true, trim=.0cm .0cm .0cm .6cm]{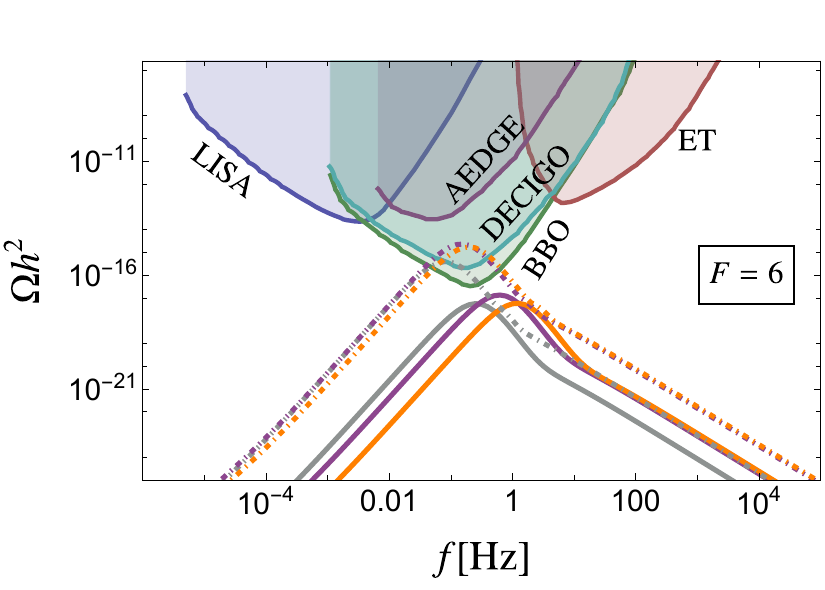} 
        \label{fig:PassFailvwN4F6}
    \end{subfigure}
    \vspace{-1.5em}
    \caption{Same as \cref{fig:PassFailVwallN3}, but for $N=4$.}
    \label{fig:PassFailVwallN4}
\end{figure}

In \cref{fig:PassFailVwallN3} and \cref{fig:PassFailVwallN4} we show the maximal envelope over the power spectrum of the stochastic gravitational wave background for different numbers of colors and flavors but fixed $\xi=1$, using different methods to calculate the wall velocity as described in \cref{sec:PT}. For the LTE or LCDF methods, we only include points where the wall velocity can be computed reliably \cite{Ai:2023see}. We also show all points which nucleated for $v_w=1$. The envelope is shown for all points, including those that fail our EFT validity criteria of unitarity and perturbativity as detailed in \cref{sec:pert}, indicated by the dot-dashed lines. The envelopes over the subset of points that pass are shown as solid lines.

\begin{figure}[h!]
    \centering
    \begin{subfigure}[b]{0.48\textwidth}
        \centering
        \includegraphics[width=\textwidth, clip=true, trim=.0cm .0cm .0cm .6cm]{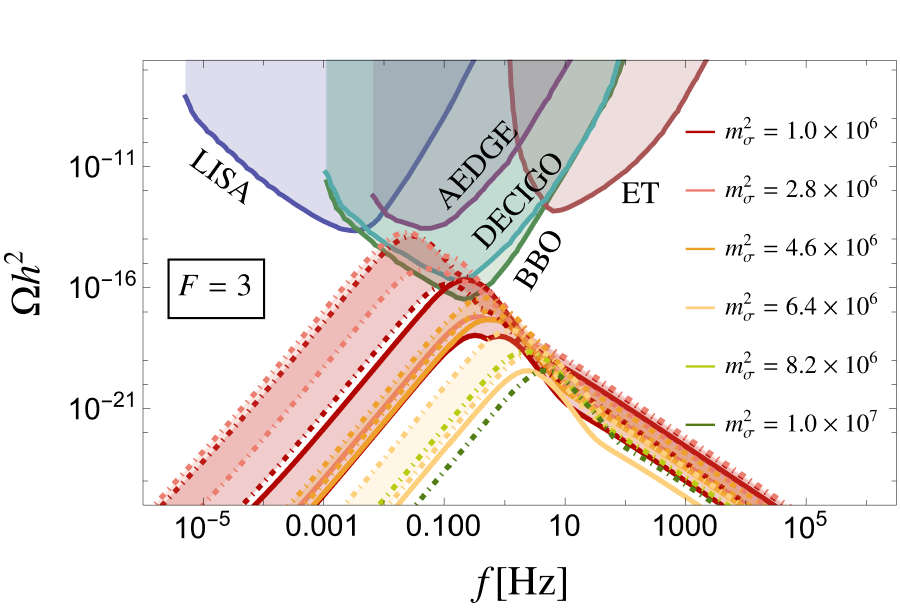} 
        \label{fig:m2SigEnvN3F3}
    \end{subfigure}
    \begin{subfigure}[b]{0.48\textwidth}
        \centering
        \includegraphics[width=\textwidth, clip=true, trim=.cm .0cm .0cm .6cm]{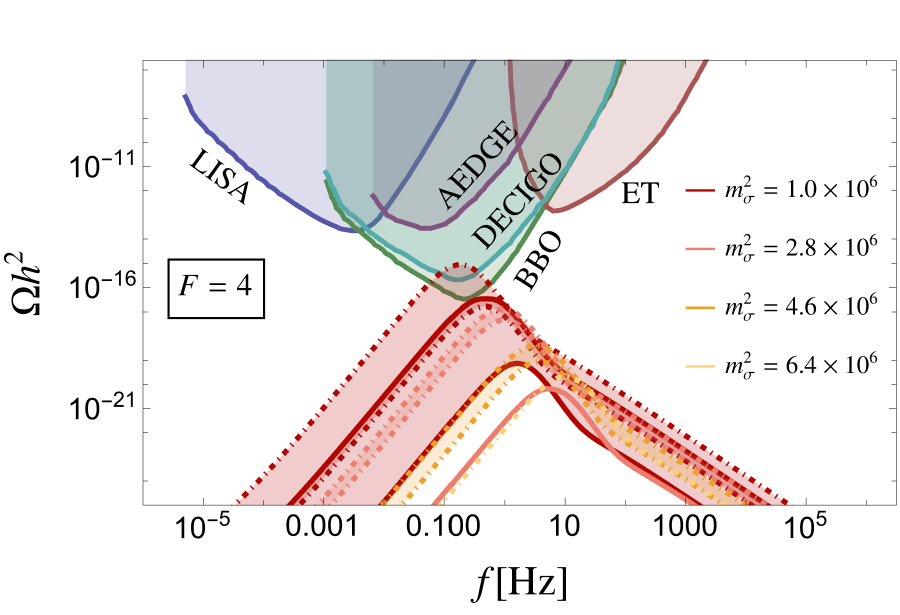} 
        \label{fig:m2SigEnvN3F4}
    \end{subfigure}
    \\
    \vspace{-1.5em}
    \begin{subfigure}[b]{0.48\textwidth}
        \centering
        \includegraphics[width=\textwidth, clip=true, trim=.0cm .0cm .0cm .6cm]{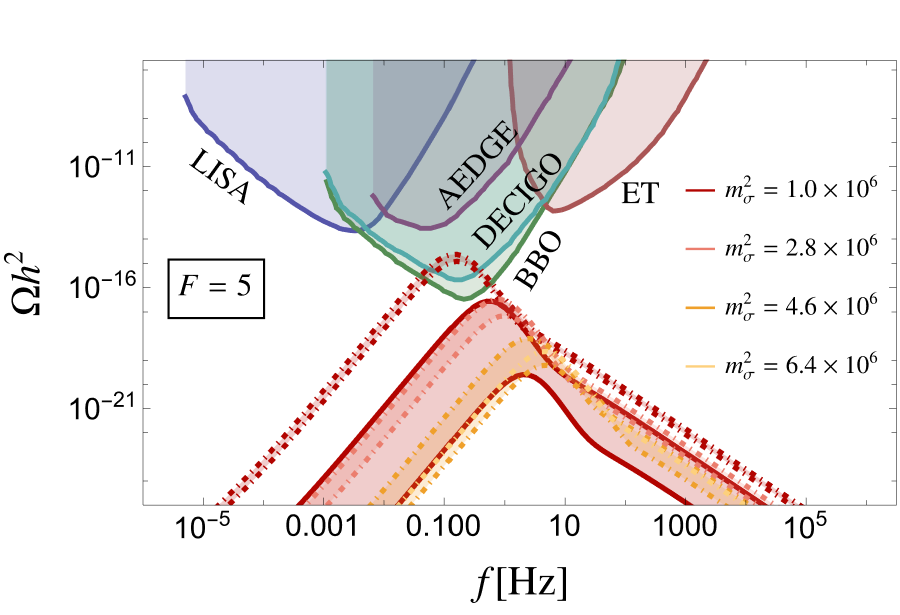} 
        \label{fig:m2SigEnvN3F5}
    \end{subfigure}
    \begin{subfigure}[b]{0.48\textwidth}
        \centering
        \includegraphics[width=\textwidth, clip=true, trim=.0cm .0cm .0cm .6cm]{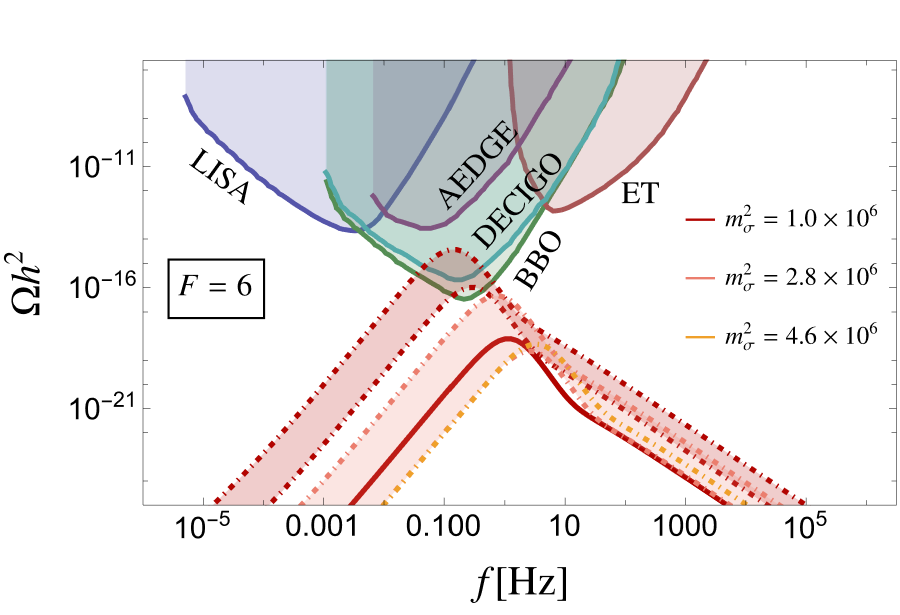} 
        \label{fig:m2SigEnvN3F6}
    \end{subfigure}
    \vspace{-1.5em}
    \caption{Gravitational wave signal strength with $m_\sigma^2$ for $N=3$. The LTE method is used to calculate $v_w$. The envelope maximizes over $\xi=1,2\;\text{and\;}5$ with no Polyakov loop contributions. When there is only a single line rather than an envelope, only one point we tested nucleated. The dot-dashed lines include all tested points. The solid lines only include points for which both the potential is unitary and our perturbativity criteria hold.}
    \label{fig:m2Sig_xiEnv_N3}
\end{figure}

\begin{figure}[h!]
    \centering
    \begin{subfigure}[b]{0.48\textwidth}
        \centering
        \includegraphics[width=\textwidth, clip=true, trim=.0cm .0cm .0cm .6cm]{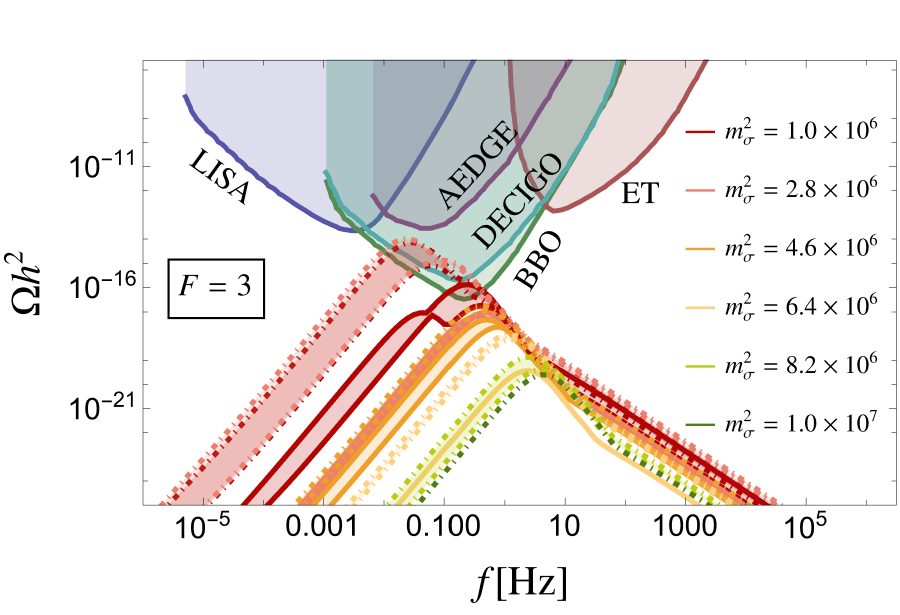} 
        \label{fig:m2SigEnvN4F3}
    \end{subfigure}
    \begin{subfigure}[b]{0.48\textwidth}
        \centering
        \includegraphics[width=\textwidth, clip=true, trim=.cm .0cm .0cm .6cm]{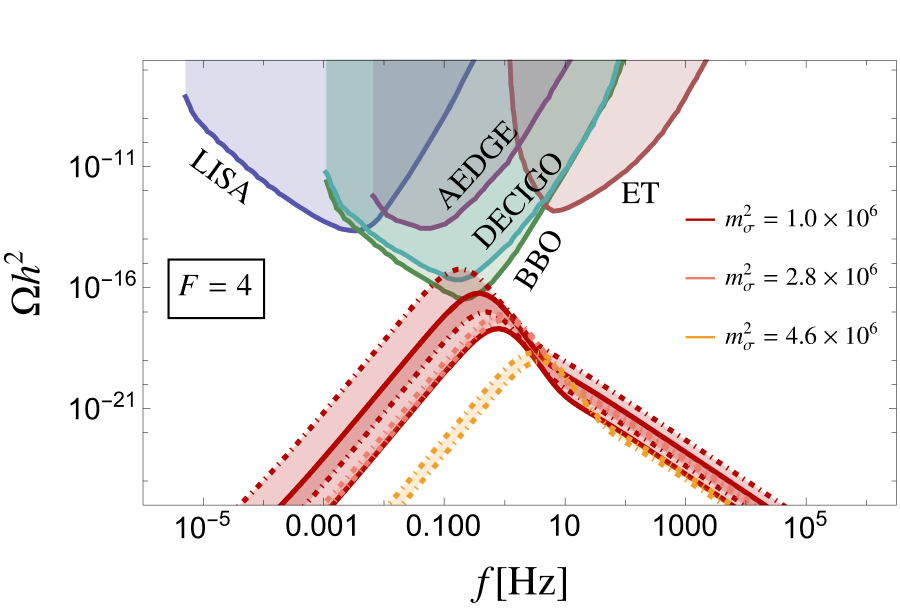} 
        \label{fig:m2SigEnvN4F4}
    \end{subfigure}
    \\
    \vspace{-1.5em}
    \begin{subfigure}[b]{0.48\textwidth}
        \centering
        \includegraphics[width=\textwidth, clip=true, trim=.0cm .0cm .0cm .6cm]{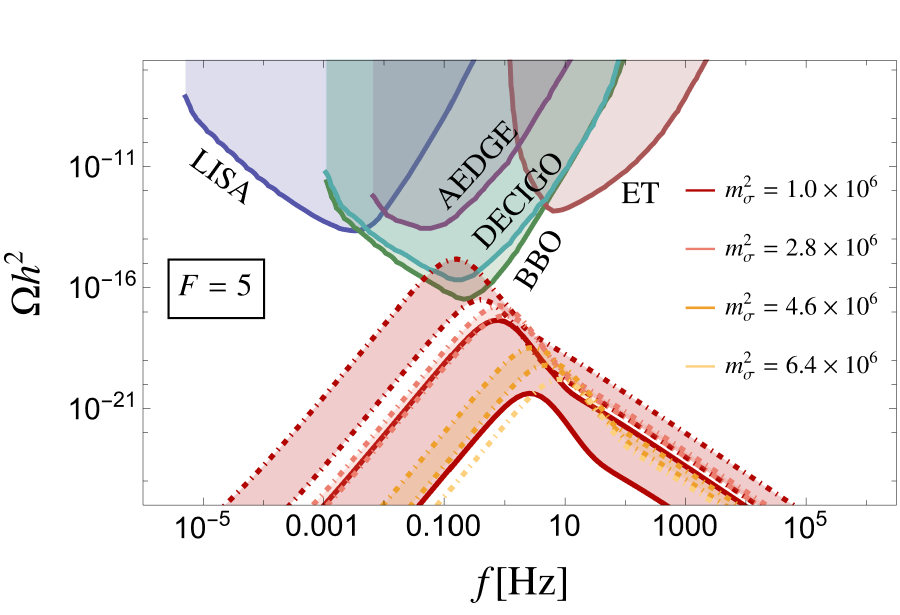} 
        \label{fig:m2SigEnvN4F5}
    \end{subfigure}
    \begin{subfigure}[b]{0.48\textwidth}
        \centering
        \includegraphics[width=\textwidth, clip=true, trim=.0cm .0cm .0cm .6cm]{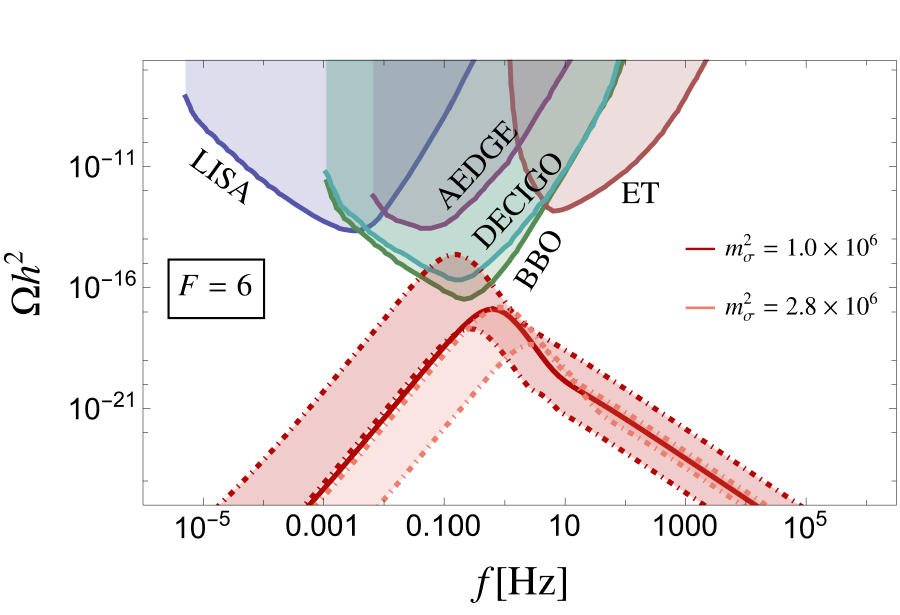} 
        \label{fig:m2SigEnvN4F6}
    \end{subfigure}
    \vspace{-1.5em}
    \caption{Same as \cref{fig:m2Sig_xiEnv_N3} with $N=4$.}
    \label{fig:m2Sig_xiEnv_N4}
\end{figure}

In \cref{fig:m2Sig_xiEnv_N3} and \cref{fig:m2Sig_xiEnv_N4}, we show the maximal envelope over different choices of $m_\sigma^2$ for different numbers of colors and flavors. The shaded regions are over the maximal signatures for $\xi=1,2\;\text{and}\;5$, and the case without the Polyakov loop improvement. This region indicates to some level the uncertainty relating to precisely the interplay between the confining and the chiral phase transitions. The wall velocity is fixed to the LTE computation as we expect this to be an optimistic method, i.e., the true wall velocity is expected to be less than this value, so these plots show the maximum reasonable power one could expect to reach from the phase transition. As noticed by \cite{Pasechnik:2023hwv}, reducing $m_\sigma$ relative to $f_\pi$ tends to yield a stronger signature. We note here that for generic potential parameters, however, $m_\sigma \sim f_\pi$, and so these signals result from somewhat tuned choices of $m_\Phi^2$ and $\lambda_{\sigma, a}$.

\begin{figure}[h!]
\centering
\begin{subfigure}[b]{0.48\textwidth}
    \centering
\includegraphics[width=\textwidth]{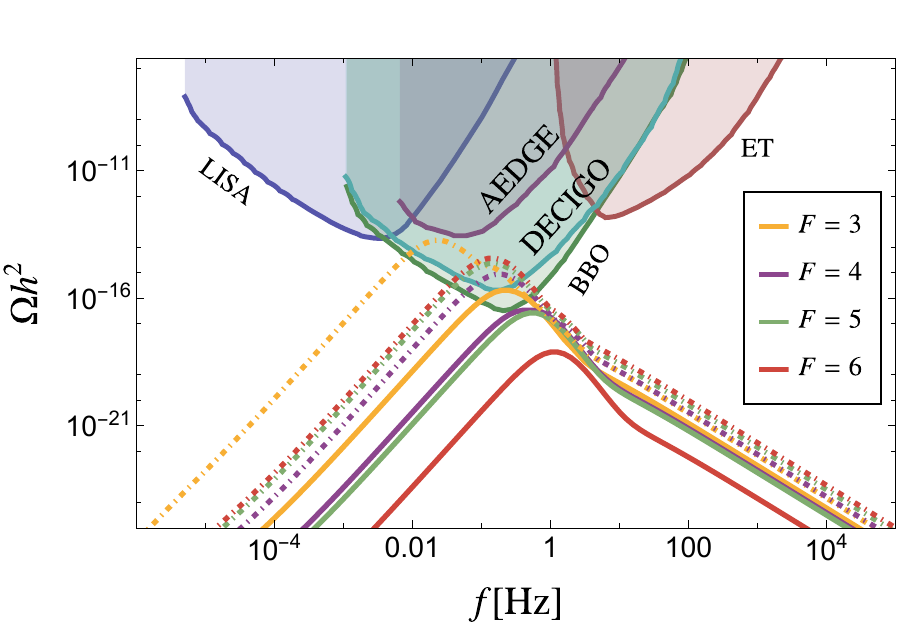}
\caption{$N=3.$}
\label{fig:FEnvN3}
\end{subfigure}
\begin{subfigure}[b]{0.48\textwidth}
        \centering
        \includegraphics[width=\textwidth, clip=true, trim=.0cm .0cm .0cm .45cm]{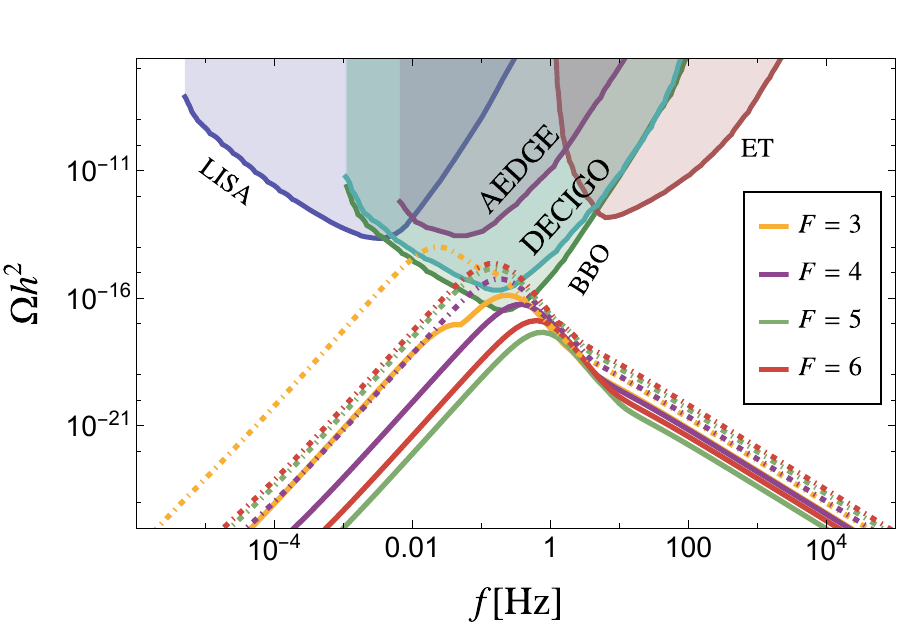} 
        \caption{$N=4$.}
        \label{fig:FEnvN4}
    \end{subfigure}
\caption{Maximal envelope plot of GW signature for various flavor benchmarks. The wall velocity is set to the LTE value, and maximization is done over the mass ranges, $\xi=1,2,\;\text{and}\;5$ and the case without Polyakov loop contributions. In \cref{fig:FEnvN3} we give the results for $N=3$ and in \cref{fig:FEnvN4} for $N=4$. The dot-dashed lines include all tested points. The solid lines only include those points for which the potential is unitary, and our perturbativity criteria hold.}
\label{fig:MaxFlavorEnvelope}
\end{figure}

Finally in \cref{fig:MaxFlavorEnvelope}, we show the maximal envelope over all tested parameter ranges, as well as $\xi=1,2,\;\text{and}\;5$ and the case without Polyakov loops. In \cref{fig:FEnvN3}, we show the maximal signal strength for $F=3,4,5\;\text{and}\;6$ for $N=3$, and \cref{fig:FEnvN4} for $N=4$. We see the maximal signal strength is highest for $F=3$, roughly decreasing with the number of flavors. Once the $U(1)_A$ breaking operator has dimension greater than four, boundedness from below constraints restrict the coupling of this term to be $c \ll f_\pi^{4-F}$. The maximal signature becomes roughly independent of the number of flavors (aside from suppression due to $g_*$). Note that we are far from the conformal window and well into the asymptotically free region for all benchmark values of $N, F$ chosen in this work.

Overall, imposing the EFT validity criteria significantly reduces the parameter space that yields observable GW signals, typically pushing the viable points to the edge of the projected reach of proposed future detectors.

\clearpage
\section{Discussion and Conclusions}
\label{sec:DiscussionConclusions}

In this work, we examined how enforcing self-consistency requirements constrains the effective description of confinement-induced phase transitions in dark non-Abelian gauge theories. Our analysis builds on a finite-temperature effective potential constructed from the LSM, augmented with Polyakov-loop potentials to incorporate confinement effects in $V_{\rm PLM}$ and medium interactions in $V_{\rm med}$, together with the CJT resummation to capture thermal effects. 

We extend earlier analyses by imposing theoretical self-consistency conditions: perturbativity, unitarity, and boundedness from below, identifying regions of parameter space in which the EFT remains trustworthy during the phase transition. We also include the $V_{\rm med}$ corrections for both $N=3$ and $N=4$. Finally, we incorporate updated estimates of $v_w$, evaluated within both the LTE and LCDF approaches. Altogether, these improvements yield a more controlled and internally consistent description of the phase transition and its resulting GW signal.

Including Polyakov-loop effects allows us to capture confinement physics more reliably at finite temperature. By extending the Polyakov-loop–improved medium effects from $N=3$ to $N=4$, we can quantify how the number of dark colors influences the transition dynamics. While the shift to $N=4$ only modestly impacts the GW signal, this extension enables a more robust exploration of $SU(N)$ confining theories beyond the well-studied $SU(3)$ case. This is particularly relevant in scenarios where low-energy EFT operators carry explicit $N$-dependence, such as those proposed in~\cite{Csaki:2023yas}. The $N=4$ results here support a more detailed examination of such operators, see analysis to appear in~\cite{Houtz:etaPrimeGW}.

Our self-consistency criteria lead to several quantitative insights into which regions of parameter space remain theoretically viable. Perturbativity provides the most restrictive condition: regions that naively predict strong phase transitions often correspond to regimes where the thermal masses of the $\sigma$ or $\pi^a$ become too small for the CJT-resummed expansion to remain reliable. Unitarity imposes a complementary constraint by excluding directions with excessively large quartic couplings, even when perturbativity alone might allow them.  
Vacuum stability depends on the size and sign of the determinant coefficient $c$. For $F\geq4$, the $\det\Phi$ operator generates $\sigma^F$ terms that can dominate the potential at high $\sigma$. The bounce solution cannot be reliably computed in this case, and so such points never entered the dataset presented in our results. Taken together, these  conditions eliminate a substantial portion of the parameter space that would otherwise appear to support a FOPT with detectable GW signals.

Even with a self-consistent EFT, the EFT may not fully capture the dynamics of a theory near confinement. The problem of a breakdown of perturbation theory when gluonic interactions become strongly coupled remains. We use methods on par with previous studies of GWs near confinement~\cite{Pasechnik:2023hwv,Reichert:2021cvs, Helmboldt:2019pan, Huang:2020crf} to account for this by incorporating existing lattice input,  but there is much room for improvement in our understanding of strongly coupled theories. Whether the parameter space where the EFT is self-consistent overlaps with the region of parameter space where the EFT more accurately captures confining dynamics is an interesting question. We leave a more detailed study to incorporate EFT improvements beyond what was done here as a problem for future work. Still, the EFT must at least be self-consistent if it has any hope of capturing the underlying physics reliably.

After imposing the EFT requirements described above, a nontrivial portion of the parameter space continues to support FOPTs capable of sourcing observable GWs, albeit observable only in the most sensitive of proposed future detectors. Enforcing these conditions generally reduces the strength of the transition relative to naive treatments, leading to a suppression of the peak amplitude and sometimes a modest shift in the peak frequency. Even so, several benchmarks yield spectra that remain within reach. In particular, the surviving signals are best probed by BBO, which retains sensitivity to the most robust predictions of the confinement-induced transitions calculated here.

Our analysis also has several limitations that could be improved upon in future work. First, the CJT potential is implemented in the Hartree–Fock approximation, which omits a subset of 2PI topologies; although we partially estimated their impact on the $\lambda_{\rm eff}$ perturbativity criteria, a complete treatment would require a systematic inclusion of the missing diagrams. Second, 
the inclusion of the SM degrees of freedom in the plasma reduces the relative drop in degrees of freedom at the phase transition, making the LCDF framework for estimating $v_w$ less reliable in this particular setting. Finally, our confinement sector depends on lattice-determined coefficients in the PLM. Extending this approach to larger $N$ or different representations would require additional nonperturbative input.

Looking ahead, our improvements to this framework can be extended to a broader class of confining theories and to concrete dark-sector constructions in which explicit chiral symmetry breaking effects play a central role~\cite{Croon:2019iuh, Houtz:etaPrimeGW}. A more refined understanding of the medium and thermal effects would further sharpen the quantitative predictions. More broadly, while enforcing EFT consistency requirements tends to suppress the predicted GW signals from confinement, the surviving observable parameter space is now supported by a more controlled and reliable EFT.

\section*{Acknowledgements}
\noindent
The authors would like to thank, Rodrigo Alonso, Djuna Croon, and James Ingoldby for helpful comments on the manuscript, and Yannick Ulrich for useful discussions. 
RH, MW, and MU are supported by the Institute for Fundamental Theory at the University of Florida. 
MU is supported by a scholarship from
Consejo Nacional de Ciencia Humanidades y Tecnologia
(CONAHCYT) of Mexico.

\bibliographystyle{JHEP}
\bibliography{Library.bib}

\clearpage
\appendix

\section{Expressions for 2$\to$2 Scattering}
\label{app:4pt-feyn}

We provide in this Appendix various useful expressions for the $2\to2$ scattering for mesons. For each scattering diagram, in Tables~\ref{tab:22scat} and \ref{tab:22scat-2}, we list:
    \begin{itemize}
        \item The four-point couplings $\lambda_\text{vev}$ obtained from expanding about the vev, $\sigma = f_\pi$. This expression is used in evaluating the unitarity constraint in~\cref{sec:unit}.
        \item The four-point couplings  $\lambda_\text{sym}$ obtained from expanding around the symmetric point with a constant background field. This expression is used in the CJT formalism to obtain all the 2PI diagrams retained after taking the Hartree Fock approximation in ~\cref{sec:CJT}. 
        \item The effective four-point couplings $\lambda_{\rm eff}$ in the finite-$T$ theory. This coupling gets large as bosonic mass shrinks in the IR, and is used to evaluate perturbativity of the expansion during the phase transition in ~\cref{sec:pert}.
    \end{itemize}
Note that in the case of expanding about the vev, we also must calculate the couplings $\kappa_\text{vev}$ resulting from 3-point interactions that result from vev insertions. These are provided in Table~\ref{tab2:3point-sigma-eta}. 

\clearpage

\begin{sidewaystable}
    \centering
 {\small
    \begin{tabular}{c|c|c|c}
    2$\to$2 process
         &  Expanding around $\sigma = f_\pi$
            & \( \lambda_{\rm sym}^{ab} \equiv \)Expanding around $\sigma = 0$ 
            & $\lambda_{\rm eff}$ \\    
    \hline
	\raisebox{-1cm}{\begin{tikzpicture}
		\draw [dashed]  (-1/1.414,1/1.414) node[anchor=south east]{$\sigma$}-- (0,0) ;
		\draw [dashed] (-1/1.414,-1/1.414) node[anchor=north east]{$\sigma$} -- (0,0);
		\draw [dashed]  (0,0)-- (1/1.414,1/1.414) node[anchor=south west]{$\sigma$};
		\draw [dashed]  (0,0)-- (1/1.414,-1/1.414) node[anchor=north west]{$\sigma$};
		\end{tikzpicture}}
            & $3\lambda_\sigma-\dfrac{c}{F^2}f_\pi^{F-4}(F)(F-1)(F-2)(F-3) $
                & $ 3 \lambda_\sigma -  \dfrac{c}{F^2} (F)(F-1)(F-2)(F- 3) \delta_{F,4} $
                    & $ \lambda_{\rm sym}^{\sigma\sigma} \,\, T \, \Bigg[\dfrac{1}{M_\sigma(\sigma,T)} \Bigg]$
                        \\
    \raisebox{-1cm}{\begin{tikzpicture}
		\draw [dashed]  (-1/1.414,1/1.414) node[anchor=south east]{$\eta'$}-- (0,0) ;
		\draw [dashed] (-1/1.414,-1/1.414) node[anchor=north east]{$\eta'$} -- (0,0);
		\draw [dashed]  (0,0)-- (1/1.414,1/1.414) node[anchor=south west]{$\eta'$};
		\draw [dashed]  (0,0)-- (1/1.414,-1/1.414) node[anchor=north west]{$\eta'$};
		\end{tikzpicture}}    
                & $3\lambda_\sigma-\dfrac{c}{F^2}f_\pi^{F-4}(F)(F-1)(F-2)(F-3)$
                    &  $ 3 \lambda_\sigma - \dfrac{c}{F^2} (F)(F-1)(F-2) (F- 3)  \delta_{F,4} \,\,$ & $ \lambda_{\rm sym}^{\eta'\eta'} \,\, T \,\Bigg[\dfrac{1}{M_{\eta'}(\sigma,T)} \Bigg]$
                           \\    
	\raisebox{-1cm}{\begin{tikzpicture}
		\draw [dashed]  (-1/1.414,1/1.414) node[anchor=south east]{$\eta'$}-- (0,0) ;
		\draw [dashed] (-1/1.414,-1/1.414) node[anchor=north east]{$\eta'$} -- (0,0);
		\draw [dashed]  (0,0)-- (1/1.414,1/1.414) node[anchor=south west]{$\sigma$};
		\draw [dashed]  (0,0)-- (1/1.414,-1/1.414) node[anchor=north west]{$\sigma$};
		\end{tikzpicture}}
              & $\lambda_\sigma +\dfrac{c}{F^2}f_\pi^{F-4}(F)(F-1)(F-2)(F-3)$
                   & $  \lambda_\sigma + \dfrac{c}{F^2} (F)(F-1)(F-2)(F- 3) \delta_{F,4} $  & $ \lambda_{\rm sym}^{\eta'\sigma} \,\, T \,\, \text{Max} \Bigg[\dfrac{1}{M_{\eta'}(\sigma,T)},\dfrac{1}{M_{\sigma}(\sigma,T)} \Bigg]$
                          \\
                          
	$\dfrac{1}{F^2-1}\sum\limits_a$\raisebox{-1cm}{\begin{tikzpicture}
		\draw [dashed]  (-1/1.414,1/1.414) node[anchor=south east]{$\sigma$}-- (0,0) ;
		\draw [dashed] (-1/1.414,-1/1.414) node[anchor=north east]{$\sigma$} -- (0,0);
		\draw [dashed]  (0,0)-- (1/1.414,1/1.414) node[anchor=south west]{$X^a$};
		\draw [dashed]  (0,0)-- (1/1.414,-1/1.414) node[anchor=north west]{$X^a$};
		\end{tikzpicture}}
            &  $(\lambda_\sigma + 2 \lambda_a) +\frac{c}{F}f_\pi^{F-4}(F-2)(F-3)$
                & $ (\lambda_\sigma + 2 \lambda_a) + \dfrac{c}{F}(F-2) (F-3) \delta_{F,4} $ &   $ \lambda_{\rm sym}^{\sigma X} \,\, T \,\, \text{Max} \Bigg[\dfrac{1}{M_\sigma(\sigma,T)},\dfrac{F^2-1}{M_{X}(\sigma,T)} \Bigg]$   \\
     $\dfrac{1}{F^2-1}\sum\limits_a$\raisebox{-1cm}{\begin{tikzpicture}
            \draw [dashed]  (-1/1.414,1/1.414) node[anchor=south east]{$\sigma$}-- (0,0) ;
            \draw [dashed] (-1/1.414,-1/1.414) node[anchor=north east]{$\sigma$} -- (0,0);
            \draw [dashed]  (0,0)-- (1/1.414,1/1.414) node[anchor=south west]{$\pi^a$};
            \draw [dashed]  (0,0)-- (1/1.414,-1/1.414) node[anchor=north west]{$\pi^a$};
        \end{tikzpicture}}
            & $\lambda_\sigma-\frac{c}{F}f_\pi^{F-4}(F-2)(F-3)$
            &  $\lambda_\sigma - \dfrac{c}{F}(F-2) (F-3) \delta_{F,4}$ & $ \lambda_{\rm sym}^{\sigma \pi} \,\, T \,\, \text{Max} \Bigg[\dfrac{1}{M_\sigma(\sigma,T)},\dfrac{F^2-1}{M_{\pi}(\sigma,T)} \Bigg]$
            \\
        $\dfrac{1}{F^2-1}\sum\limits_a$\raisebox{-1cm}{\begin{tikzpicture}
            \draw [dashed]  (-1/1.414,1/1.414) node[anchor=south east]{$\eta'$}-- (0,0) ;
            \draw [dashed] (-1/1.414,-1/1.414) node[anchor=north east]{$\eta'$} -- (0,0);
            \draw [dashed]  (0,0)-- (1/1.414,1/1.414) node[anchor=south west]{$X^a$};
            \draw [dashed]  (0,0)-- (1/1.414,-1/1.414) node[anchor=north west]{$X^a$};
        \end{tikzpicture}}
            & $\lambda_\sigma-\dfrac{c}{F}f_\pi^{F-4}(F-2)(F-3)$
            & $\lambda_\sigma - \dfrac{c}{F} (F-2) (F-3) \delta_{ F, 4}$   &
            $ \lambda_{\rm sym}^{\eta' X} \,\, T \,\, \text{Max} \Bigg[\dfrac{1}{M_{\eta'}(\sigma,T)},\dfrac{F^2-1}{M_{X}(\sigma,T)} \Bigg]$
            \\
            
    \end{tabular}
    }
    \caption{
    The Feynman rules for 2$\to$2 scattering of mesons in the broken phase (second column), the symmetric phase (third column), as well as the resulting effective coupling in the thermal corrections (last column), we define the symmetrized quartic couplings
\(
  \lambda_{\text{sym}}^{ab}\) with
  a,b  \( \in \{\sigma,\eta,X,\pi\}\). These expressions are used to derive the unitarity constraint, the CJT corrections, and the perturbativity constraint, respectively. This list continues in Table~\ref{tab:22scat-2}. 
    }  \label{tab:22scat}
\end{sidewaystable}

\clearpage

\begin{sidewaystable}
    \centering
    {\small
    \begin{tabular}{c|c|c|c}
        2$\to$2 process
         &  Expand around $\sigma = f_\pi$
         &  \( \lambda_{\rm sym}^{ab} \equiv \)Expanding around $\sigma = 0$ 
         &  $\lambda_{\rm eff}$ \\    
        \hline
       
        $\dfrac{1}{F^2-1}\sum\limits_{a}$\raisebox{-1cm}{\begin{tikzpicture}
            \draw [dashed]  (-1/1.414,1/1.414) node[anchor=south east]{$\eta'$}-- (0,0) ;
            \draw [dashed] (-1/1.414,-1/1.414) node[anchor=north east]{$\eta'$} -- (0,0);
            \draw [dashed]  (0,0)-- (1/1.414,1/1.414) node[anchor=south west]{$\pi^a$};
            \draw [dashed]  (0,0)-- (1/1.414,-1/1.414) node[anchor=north west]{$\pi^a$};
        \end{tikzpicture}}
            & $(\lambda_\sigma + 2 \lambda_a)+ \dfrac{c}{F}f_\pi^{F-4}(F-2)(F-3)$
            & $(\lambda_\sigma + 2 \lambda_a) + \dfrac{c}{F} (F-2) (F-3) \delta_{F, 4}$  & $ \lambda_{\rm sym}^{\eta' \pi} \,\, T \,\,\text{Max} \Bigg[\dfrac{1}{M_{\eta'}(\sigma,T)},\dfrac{F^2-1}{M_{\pi}(\sigma,T)} \Bigg]$
            \\
       $\dfrac{1}{(F^2-1)^2}\sum\limits_{a,b}$ \raisebox{-1cm}{\begin{tikzpicture}
            \draw [dashed]  (-1/1.414,1/1.414) node[anchor=south east]{$X^b$}-- (0,0) ;
            \draw [dashed] (-1/1.414,-1/1.414) node[anchor=north east]{$X^b$} -- (0,0);
            \draw [dashed]  (0,0)-- (1/1.414,1/1.414) node[anchor=south west]{$X^a$};
            \draw [dashed]  (0,0)-- (1/1.414,-1/1.414) node[anchor=north west]{$X^a$};
        \end{tikzpicture}}
            & $  \begin{array}{r}
                \lambda \sigma \dfrac{F^2+1}{F^2-1} + \lambda_a \dfrac{F^2-4}{F^2-1} + \\
                \dfrac{c}{F}f_\pi^{F-4}  \dfrac{(F+1)(F-2)(F-3)}{ F^2 - 1}
            \end{array} $
            & $ \begin{array}{r}
                \lambda_\sigma \dfrac{F^2+1}{F^2-1} + \lambda_a \dfrac{F^2-4}{F^2-1} + \\
                \dfrac{c}{F}  \dfrac{(F+1)(F-2)(F-3)}{ F^2 - 1} \delta_{F, 4}
            \end{array}$ & $ \lambda_{\rm sym}^{XX} \,\, T \,\Bigg[\dfrac{F^2-1}{M_{X}(\sigma,T)} \Bigg]$
            \\
        $\dfrac{1}{(F^2-1)^2}\sum\limits_{a,b}$\raisebox{-1cm}{\begin{tikzpicture}
            \draw [dashed]  (-1/1.414,1/1.414) node[anchor=south east]{$X^a$}-- (0,0) ;
            \draw [dashed] (-1/1.414,-1/1.414) node[anchor=north east]{$X^a$} -- (0,0);
            \draw [dashed]  (0,0)-- (1/1.414,1/1.414) node[anchor=south west]{$\pi^b$};
            \draw [dashed]  (0,0)-- (1/1.414,-1/1.414) node[anchor=north west]{$\pi^b$};
        \end{tikzpicture}}
          & $\begin{array}{r}
          \lambda_\sigma + \lambda_a \dfrac{F^2}{F^2-1}\\
            -\dfrac{c}{F}f_\pi^{F-4} \dfrac{(F+1)(F-2)(F-3)}{F^2 -1}
            \end{array} $
          & $\begin{array}{r}
            \lambda_\sigma + \lambda_a \dfrac{F^2}{F^2-1} \\
            - \dfrac{c}{F} \dfrac{(F+1)(F-2)(F-3)}{F^2 -1} \delta_{F, 4}
            \end{array}$ & $ \lambda_{\rm sym}^{X\pi} \,\, T \,\, \text{Max} \Bigg[\dfrac{F^2-1}{M_{X}(\sigma,T)},\dfrac{F^2-1}{M_{\pi}(\sigma,T)} \Bigg]$
          \\
        $\dfrac{1}{(F^2-1)^2}\sum\limits_{a,b}$\raisebox{-1cm}{\begin{tikzpicture}
            \draw [dashed]  (-1/1.414,1/1.414) node[anchor=south east]{$\pi^a$}-- (0,0) ;
            \draw [dashed] (-1/1.414,-1/1.414) node[anchor=north east]{$\pi^a$} -- (0,0);
            \draw [dashed]  (0,0)-- (1/1.414,1/1.414) node[anchor=south west]{$\pi^b$};
            \draw [dashed]  (0,0)-- (1/1.414,-1/1.414) node[anchor=north west]{$\pi^b$};
        \end{tikzpicture}}
          & $\begin{array}{r}
          \lambda_\sigma \dfrac{F^2+1}{F^2-1} + \lambda_a\dfrac{F^2-4}{F^2-1} \\
            +\dfrac{c}{F}f_\pi^{F-4}\dfrac{ (F+1)(F-2)(F-3)}{F^2 -1}
            \end{array} $
          & $\begin{array}{r}
            \lambda_\sigma \dfrac{F^2+1}{F^2-1} + \lambda_a\dfrac{F^2-4}{F^2-1} \\
            + \dfrac{c}{F} \dfrac{ (F+1)(F-2)(F-3)}{F^2 -1} \delta_{F,4}
            \end{array}$ & $ \lambda_{\rm sym}^{\pi\pi} \,\, T \,\Bigg[\dfrac{F^2-1}{M_{\pi}(\sigma,T)}\Bigg]$
    \end{tabular}
    }
    \caption{(Continuation of Table~\ref{tab:22scat}) The Feynman rules for 2$\to$2 scattering of mesons in the broken phase (second column), the symmetric phase (third column), as well as the resulting effective coupling in the thermal corrections (last column),  we define the symmetrized quartic couplings
\(
  \lambda_{\text{sym}}^{ab}\) with
  a,b  \( \in \{\sigma,\eta,X,\pi\}\). These expressions are used to derive the unitarity constraint, the CJT corrections, and the perturbativity constraint, respectively.}
    \label{tab:22scat-2}
\end{sidewaystable}

\clearpage

\begin{sidewaystable}
    \centering
    \begin{tabular}{c|c|c}
        3--point vertex
         & Expand around $\sigma = f_\pi$
         & Expand around $\sigma = 0$ \\ \hline
\raisebox{-1.0cm}{%
\begin{tikzpicture}
  \coordinate (v) at (0,0);
  \draw[dashed] (-1.4,1.0) node[left] {$\sigma$} -- (v);
  \draw[dashed] (-1.4,-1.0) node[left] {$\sigma$} -- (v);
  \draw[dashed] (v) -- (1.4,0) node[right] {$\sigma$};
\end{tikzpicture}}
 & $\,3\lambda_\sigma f_\pi
           - \dfrac{c}{F}\,f_\pi^{F-3}\,(F-1)(F-2)$
 & $-  \dfrac{c}{F} (F-1) (F-2) \,\,\delta_{F,3}$
 \\[2.2ex]
\raisebox{-1.0cm}{%
\begin{tikzpicture}
  \coordinate (v) at (0,0);
  \draw[dashed] (-1.4,1.0) node[left] {$\sigma$} -- (v);
  \draw[dashed] (-1.4,-1.0) node[left] {$\eta'$} -- (v);
  \draw[dashed] (v) -- (1.4,0) node[right] {$\sigma$};
\end{tikzpicture}}
 & $0$
 & $0$
 \\[2.2ex]
\raisebox{-1.0cm}{%
\begin{tikzpicture}
  \coordinate (v) at (0,0);
  \draw[dashed] (-1.4,1.0) node[left] {$\eta'$} -- (v);
  \draw[dashed] (-1.4,-1.0) node[left] {$\eta'$} -- (v);
  \draw[dashed] (v) -- (1.4,0) node[right] {$\sigma$};
\end{tikzpicture}}
 & $\,\lambda_\sigma f_\pi
           + \dfrac{c}{F}\,f_\pi^{F-3}\,(F-1)(F-2)$
 & $\dfrac{c}{F} (F-1) (F-2)\delta_{F,3} $
 \\[2.2ex]
\raisebox{-1.0cm}{%
\begin{tikzpicture}
  \coordinate (v) at (0,0);
  \draw[dashed] (-1.4,1.0) node[left] {$\eta'$} -- (v);
  \draw[dashed] (-1.4,-1.0) node[left] {$\eta'$} -- (v);
  \draw[dashed] (v) -- (1.4,0) node[right] {$\eta'$};
\end{tikzpicture}}
 & $0$
 & $0$
 \\[2.2ex]

$\dfrac{1}{F^2-1}\sum\limits_a$\raisebox{-1.0cm}{%
\begin{tikzpicture}
  \coordinate (v) at (0,0);
  \draw[dashed] (-1.4,1.0) node[left] {$\sigma$} -- (v);
  \draw[dashed] (-1.4,-1.0) node[left] {$X^a$} -- (v);
  \draw[dashed] (v) -- (1.4,0) node[right] {$X^a$};
\end{tikzpicture}}
 & $ \; (\lambda_\sigma + 2\lambda_a)f_\pi
          + \dfrac{c}{F}\,f_\pi^{F-3}\,(F-2)
  $
 & $\dfrac{c}{F}(F-2) \delta_{F,3}$
 \\[2.2ex]
$\dfrac{1}{F^2-1}\sum\limits_a$\raisebox{-1.0cm}{%
\begin{tikzpicture}
  \coordinate (v) at (0,0);
  \draw[dashed] (-1.4,1.0) node[left] {$\sigma$} -- (v);
  \draw[dashed] (-1.4,-1.0) node[left] {$\pi^a$} -- (v);
  \draw[dashed] (v) -- (1.4,0) node[right] {$X^a$};
\end{tikzpicture}}
 & $0$
 & $0$
 \\[2.2ex]
 \end{tabular}
 \caption{Trilinear meson couplings in the $\sigma$–$\eta'$ sector. The second column shows
      the Feynman rules expanded around the broken minimum $\sigma=f_\pi$, and the third column
      around the symmetric point $\sigma=0$.}
 \end{sidewaystable}

\begin{sidewaystable}
\begin{tabular}{c|c|c}
        3--point vertex
         & Expand around $\sigma = f_\pi$
         & Expand around $\sigma = 0$ \\ \hline
$\dfrac{1}{F^2-1}\sum\limits_a$\raisebox{-1.0cm}{%
\begin{tikzpicture}
  \coordinate (v) at (0,0);
  \draw[dashed] (-1.4,1.0) node[left] {$\sigma$} -- (v);
  \draw[dashed] (-1.4,-1.0) node[left] {$\pi^a$} -- (v);
  \draw[dashed] (v) -- (1.4,0) node[right] {$\pi^a$};
\end{tikzpicture}}
 & $\lambda_\sigma f_\pi
          - \dfrac{c}{F}\,f_\pi^{F-3}\,(F-2)
   $
 & $- \dfrac{c}{F}\,(F-2)\delta_{F,3}$
 \\[2.2ex]
$\dfrac{1}{F^2-1}\sum\limits_a$\raisebox{-1.0cm}{%
\begin{tikzpicture}
  \coordinate (v) at (0,0);
  \draw[dashed] (-1.4,1.0) node[left] {$\eta'$} -- (v);
  \draw[dashed] (-1.4,-1.0) node[left] {$X^a$} -- (v);
  \draw[dashed] (v) -- (1.4,0) node[right] {$X^a$};
\end{tikzpicture}}
 & $0$
 & $0$
 \\[2.2ex]

$\dfrac{1}{F^2-1}\sum\limits_a$\raisebox{-1.0cm}{%
\begin{tikzpicture}
  \coordinate (v) at (0,0);
  \draw[dashed] (-1.4,1.0) node[left] {$\eta'$} -- (v);
  \draw[dashed] (-1.4,-1.0) node[left] {$\pi^a$} -- (v);
  \draw[dashed] (v) -- (1.4,0) node[right] {$X^a$};
\end{tikzpicture}}
 & $\lambda_a f_\pi
          - \dfrac{c}{F}\,f_\pi^{F-3}\,(F-2)
 $
 & $\dfrac{c}{F}\,(F-2)\delta_{F,3}$
 \\[2.2ex]
$\dfrac{1}{F^2-1}\sum\limits_a$\raisebox{-1.0cm}{%
\begin{tikzpicture}
  \coordinate (v) at (0,0);
  \draw[dashed] (-1.4,1.0) node[left] {$\eta'$} -- (v);
  \draw[dashed] (-1.4,-1.0) node[left] {$\pi^a$} -- (v);
  \draw[dashed] (v) -- (1.4,0) node[right] {$\pi^a$};
\end{tikzpicture}}
 & $0$
 & $0$
 \\[2.2ex]
\raisebox{-1.0cm}{%
\begin{tikzpicture}
  \coordinate (v) at (0,0);
  \draw[dashed] (-1.4,1.0) node[left] {$X^a$} -- (v);
  \draw[dashed] (-1.4,-1.0) node[left] {$X^b$} -- (v);
  \draw[dashed] (v) -- (1.4,0) node[right] {$X^c$};
\end{tikzpicture}}
 & $3\lambda_a\sqrt{\dfrac{F}{2}}d^{abc}-\sqrt{\dfrac{2}{F}}\,cf_\pi^{F-3}\,d^{abc}$
 & $-\sqrt{\dfrac{2}{F}}\,c\,d^{abc}\, \delta_{F,3}$
 \\[2.2ex]
\raisebox{-1.0cm}{%
\begin{tikzpicture}
  \coordinate (v) at (0,0);
  \draw[dashed] (-1.4,1.0) node[left] {$X^a$} -- (v);
  \draw[dashed] (-1.4,-1.0) node[left] {$X^b$} -- (v);
  \draw[dashed] (v) -- (1.4,0) node[right] {$\pi^c$};
\end{tikzpicture}}
 & $0$
 & $0$
 \\[1.2ex]
    \end{tabular}
    \caption{Trilinear meson couplings in the $\sigma$–$\eta'$ sector. The second column shows
      the Feynman rules expanded around the broken minimum $\sigma=f_\pi$, and the third column
      around the symmetric point $\sigma=0$.}
    \label{tab2:3point-sigma-eta}   
\end{sidewaystable}

\begin{sidewaystable}
    \centering
    \begin{tabular}{c|c|c}
        3--point vertex
         & Expand around $\sigma = f_\pi$
         & Expand around $\sigma = 0$ \\ \hline
\raisebox{-1.0cm}{%
\begin{tikzpicture}
  \coordinate (v) at (0,0);
  \draw[dashed] (-1.4,1.0) node[left] {$X^a$} -- (v);
  \draw[dashed] (-1.4,-1.0) node[left] {$\pi^c$} -- (v);
  \draw[dashed] (v) -- (1.4,0) node[right] {$\pi^b$};
\end{tikzpicture}}
 & $-i\Big[\lambda_a\sqrt{\dfrac{F}{2}}\,d^{abc}\,+\sqrt{\dfrac{2}{F}}\,cf_\pi^{F-3}\,d^{abc}\Big]$
 & $+\sqrt{\dfrac{2}{F}}\,c\,d^{abc}\,\delta_{F,3}$
 \\[2.2ex]
\raisebox{-1.0cm}{%
\begin{tikzpicture}
  \coordinate (v) at (0,0);
  \draw[dashed] (-1.4,1.0) node[left] {$\pi^a$} -- (v);
  \draw[dashed] (-1.4,-1.0) node[left] {$\pi^b$} -- (v);
  \draw[dashed] (v) -- (1.4,0) node[right] {$\pi^c$};
\end{tikzpicture}}
 & $0$
 & $0$
 \\[1ex]

 \end{tabular}
    \caption{Trilinear meson couplings in the $\sigma$–$\eta'$ sector. The second column shows
      the Feynman rules expanded around the broken minimum $\sigma=f_\pi$, and the third column
      around the symmetric point $\sigma=0$.}
    \label{tab2:3point-sigma-eta}   
\end{sidewaystable}

\clearpage

\section{Dressed Mass Equations}
\label{app:cjt}
The CJT formalism admits the the following system of equations for the thermal dressed masses:
\begin{align}
    M^2_\sigma(\sigma,T)
        &=m^2_\sigma(\sigma)+\frac{T^2}{4\pi^2}\Big\{(3\lambda_\sigma-c_1)I_B(R_\sigma^2)+((F^2-1)(\lambda_\sigma+2\lambda_a)+c_2\tilde{c})I_B(R^2_X)
        \nonumber\\\
        &\qquad\qquad\qquad\qquad+(\lambda_\sigma+c_1)I_B(R^2_{\eta'})+((F^2-1)\lambda_\sigma-c_2\tilde{c})I_B(R^2_\pi)\Big\}
    \\
    M^2_{\eta'}(\sigma,T)
        &=m^2_{\eta'}(\sigma)+\frac{T^2}{4\pi^2}\Big\{(3\lambda_\sigma-c_1)I_B(R_{\eta'}^2)+((F^2-1)(\lambda_\sigma+2\lambda_a)+c_2\tilde{c})I_B(R_\pi^2)
        \nonumber\\
        &\qquad\qquad\qquad\qquad+(\lambda_\sigma+c_1)I_B(R^2_\sigma)+((F^2-1)\lambda_\sigma-c_2\tilde{c})I_B(R_X^2)\Big\}
        \\
    M^2_{X}(\sigma,T)
        &=m^2_{X}(\sigma)
            +\frac{T^2}{4\pi^2}\Big\{
                (\lambda_\sigma+2\lambda_a+c_2)I_B(R_\sigma^2)
        \nonumber\\
        &\qquad\qquad\qquad\qquad
                +((F^2+1)\lambda_\sigma+(F^2-4)\lambda_a-c_3)I_B(R_X^2)
        +(\lambda_\sigma-c_2)I_B(R_{\eta'}^2)
        \nonumber\\
        &\qquad\qquad\qquad\qquad
        +(\textcolor{black}{(F^2-1)}\lambda_\sigma+F^2\lambda_a+c_3)I_B(R^2_\pi)\Big\}\\
    M^2_{\pi
    }(\sigma,T)
        &=m^2_{\pi}(\sigma)
            +\frac{T^2}{4\pi^2}\Big\{
                (\lambda_\sigma+2\lambda_a+c_2)I_B(R_{\eta'}^2)
        \nonumber\\
        &\qquad\qquad\qquad\qquad
             +((F^2+1)\lambda_\sigma+(F^2-4)\lambda_a-c_3)I_B(R_\pi^2)
            +(\lambda_\sigma-c_2)I_B(R_\sigma^2)
        \nonumber\\
        &\qquad\qquad\qquad\qquad
            +(\textcolor{black}{(F^2-1)}\lambda_\sigma+F^2\lambda_a+c_3)I_B(R^2_X)\Big\} \,,
\end{align}
where
\begin{align}
    c_1&=\frac{c}{F}(F-1)(F-2)(F-3)\delta(F,4)
    \label{eq:c1}
    \\
    c_2&=\frac{c}{F}(F-2)(F-3)\delta(F,4)\\
    c_3&=\frac{c}{F}(F^3-4F^3+F+6)\delta(F,4)
    \label{eq:c3}
    \\
    \tilde{c}&=F^2-1\,.
\end{align}
And the tree-level effective masses are
\begin{align}
    m_\sigma^2(\sigma)&=-m^2-\frac{c}{F}\left(F-1
\right)\sigma^{F-2}+\frac{3\lambda_\sigma}{2}\sigma^2\\
    m_{\eta'}^2(\sigma)&=-m^2+\frac{c}{F}\left(F-1
\right)\sigma^{F-2}+\frac{\lambda_\sigma}{2}\sigma^2\\
    m_X^2(\sigma)&=-m^2+\frac{c}{F}\sigma^{F-2}+\frac{1}{2}(\lambda_\sigma+2\lambda_a)\sigma^2 \\
    m_\pi^2(\sigma)&=-m^2-\frac{c}{F}\sigma^{F-2}+\frac{\lambda_\sigma}{2}\sigma^2
\end{align}
These expressions arise from solving the Schwinger-Dyson equations and are simplified using the Hartree-Fock approximation.

\clearpage
\section{Perturbative Unitarity Bounds from Elastic Scattering}

In this appendix we detail the formulae for computing the tree-level perturbative unitarity bounds used in this work. While much of this is already detailed in the literature including in \cite{Goodsell:2018tti}, we include the explicit expressions for interest.

Using the statement of the optical theorem in terms of forward scattering,
\begin{align}
    \mathfrak{Im}\,{\mathcal{M}}(\phi_{1,a}\phi_{2,b}\rightarrow\phi_{1,a}\phi_{2,b},\theta=0)=2E_\text{CM}|\vec{p}_i|\sum_f\sigma_\text{tot}(\phi_{1,a}\phi_{2,b}\rightarrow f)
\end{align}
where $f$ denotes all final-states, and $\vec{p}_i$ denotes the 3-momentum of the incoming scattering pair in the centre of mass frame. We use the notation $\phi_{1,a}$ to represent a particle of type $1$ and mass $m_1$ and group index $a$. 

Considering the tree-level elastic scattering process provides an approximate upper bound on the matrix element. However, for the LSM considered here we have many indistinguishable species of $\pi$ and $X$ mesons which must be incorporated into the bounds. We do this by taking the average scattering over the states, e.g. 
\begin{align}
    \frac{1}{(F^2-1)^2}\sum_{a,b}\mathcal{M}(\pi_a\pi_b\rightarrow\pi_a\pi_b)\implies\mathcal{M}(\bar{\pi}\bar{\pi}\rightarrow\bar{\pi}\bar{\pi}),
\end{align}
where, for brevity, we denote the group-averaged fields e.g. $\bar{\phi}_1$. While we have provided most diagrams already appropriately averaged in \cref{tab:22scat,tab:22scat-2,tab2:3point-sigma-eta}, note that care must be taken considering the s-, t- and u-channel scattering topologies. In particular due to the 3-point vertices involving the symmetric tensor $d^{abc}$. Note that any Lagrangian terms which may appear to depend on the antisymmetric tensor $f^{abc}$ must vanish when taken to Feynman vertex form. In our normalisation,
\begin{align}
    \tr\left[T_a T_b\right]=\frac{1}{2} && \sum_c d^{abc}d^{abc}=\frac{F^2-4}{F}\delta^{ab}.
\end{align}
In terms of the barred fields, we have for the optical theorem:
\begin{align}
    \mathfrak{Im}\,\mathcal{M}(\bar{\phi}_1\bar{\phi}_2\rightarrow\bar{\phi}_1\bar{\phi}_2,\theta=0)\geq2E_\text{CM}|\vec{p}_i|\sigma_\text{tot}(\bar{\phi}_1\bar{\phi}_2\rightarrow\bar{\phi}_1\bar{\phi}_2).
\end{align}
Now performing the usual decomposition into partial waves, 
\begin{align}
    \mathcal{M}(\theta)=16\pi\sum_{j=0}^\infty a_j(2j+1)P_j(\cos\theta),
\end{align}
where $P_j$ are Legendre Polynomials normalised to 1, leads to the usual bound on the $j$-th partial wave:
\begin{align}
    \mathfrak{Im}\,a^j\leq|a^j|^2,
\end{align}
if the total cross-section is well-approximated by the elastic one. For this work, we consider the bound on the zeroth partial wave.

The scattering matrix from considering the diagram topologies in \cref{fig:DiagramTopologies} is
\begin{align}
    \mathcal{A}=-\lambda^{\bar{\phi}_1\bar{\phi}_2}_\text{vev} -\sum_i \kappa_\text{vev}^{i\bar{\phi}_1\bar{\phi}_2}\kappa_\text{vev}^{i\bar{\phi}_1\bar{\phi}_2}\frac{1}{s-m_{i,s}^2}-\sum_j\kappa_\text{vev}^{j\bar{\phi}_1\bar{\phi}_1}\kappa_\text{vev}^{j\bar{\phi}_2\bar{\phi}_2}\frac{1}{t-m_{j,t}^2}-\sum_k\kappa_\text{vev}^{k\bar{\phi}_1\bar{\phi}_2}\kappa_\text{vev}^{k\bar{\phi}_1\bar{\phi}_2}\frac{1}{u-m_{k,u}^2}.
\end{align}
The sums are over all contributing propagators in \cref{fig:DiagramTopologies}.
And finally adapting the expression from \cite{Goodsell:2018tti} for the s-dependence of the zeroth partial wave,
\begin{align}
    a_0(s)&=-\frac{2^{\delta_{\phi_1,\phi_2}}}{16\pi}\Bigg[\sqrt{\lambda(s,m_1^2,m_2^2)}\left(\lambda_\text{vev}^{\phi_1\phi_2}+\sum_i \kappa_\text{vev}^{i\bar{\phi}_1\bar{\phi}_2}\kappa_\text{vev}^{i\bar{\phi}_1\bar{\phi}_2}\frac{1}{s-m_{i,s}^2}\right)\\
    &\qquad\qquad-\sum_j\kappa_\text{vev}^{j\bar{\phi}_1\bar{\phi}_1}\kappa_\text{vev}^{j\bar{\phi}_2\bar{\phi}_2}f^j_{t,\text{el}}(s,m_1^2,m_2^2)-\sum_k\kappa_\text{vev}^{k\bar{\phi}_1\bar{\phi}_2}\kappa_\text{vev}^{k\bar{\phi}_1\bar{\phi}_2}f^k_{u,\text{el}}(s,m_1^2,m_2^2)\Bigg]
\end{align}
where
\begin{align}
    \lambda(s,m_1^2,m_2^2)=\frac{1}{s^2}\left(s^2+m_1^4+m_2^4-2m_1^2m_2^2-2sm_1^2-2sm_2^2\right)
\end{align}
and
\begin{align}
    f^j_{t,\text{el}}&=\frac{1}{s}\frac{1}{\sqrt{\lambda(s,m_1^2,m_2^2)}}\log\left[\frac{m_{j,t}^2s}{(m_1^2-m_2^2)^2-2(m_1^2+m_2^2)s+m_{j,t}^2s+s^2}\right]\\
    f^k_{u,\text{el}}&=\frac{1}{s}\frac{1}{\sqrt{\lambda(s,m_1^2,m_2^2)}}\log\left[\frac{(m_1^2-m_2^2)^2-m_{k,u}^2s}{s(2m_1^2+2m_2^2-m_{k,u}^2-s)}\right].
\end{align}
The s-,t-, and u-poles which appear in the amplitude we detail in the main text in \cref{sec:unit}.



\end{document}